\documentclass[reprint,aps,prb,floatfix]{revtex4-2}
 
\usepackage{graphicx}% Include figure files
\usepackage{dcolumn}% Align table columns on decimal point
\usepackage{bm}% bold math
\usepackage{amsmath, amsthm, amssymb, amsfonts, textcomp, gensymb, subfigure, import, color, siunitx, float, bm, graphicx, url, comment}

\begin{document}

\preprint{APS/123-QED}

\title{Low temperature state in strontium titanate microcrystals using \textit{in situ} multi-reflection Bragg coherent x-ray diffraction imaging}% Force line breaks with \\
%\thanks{A footnote to the article title}%

\author{David Yang}
\email{dyang2@bnl.gov}
\affiliation{Condensed Matter Physics and Materials Science Department, Brookhaven National Laboratory, Upton, NY 11973, USA}

\author{Sung Soo Ha}%
%\email{sungsoo@sogang.ac.kr}
\affiliation{Center for Ultrafast Phase Transformation, Department of Physics, Sogang University, Seoul 04107, Republic of Korea}

\author{Sungwook Choi}
%\email{hope9784@naver.com}
\affiliation{Center for Ultrafast Phase Transformation, Department of Physics, Sogang University, Seoul 04107, Republic of Korea}

\author{Jialun Liu}
%\email{jialun.liu.17@ucl.ac.uk}
\affiliation{London Centre for Nanotechnology, University College London, London WC1E 6BT, UK}

\author{Daniel Treuherz}
%\email{daniel.treuherz@hotmail.com}
%\affiliation{Department of Physics, University of Cambridge, Cambridge CB2 1TN, UK}
\affiliation{London Centre for Nanotechnology, University College London, London WC1E 6BT, UK}

\author{Nan Zhang}
%\email{nzhang1@xjtu.edu.cn}
\affiliation{Electronic Materials Research Laboratory, Key Laboratory of the Ministry of Education \& International Center for Dielectric Research,School of Electronic Science and Engineering, Xi’an Jiaotong University, Xi’an 710049, China}

\author{Zheyi An}
%\email{anzheyi@xjtu.edu.cn}
\affiliation{Electronic Materials Research Laboratory, Key Laboratory of the Ministry of Education \& International Center for Dielectric Research,School of Electronic Science and Engineering, Xi’an Jiaotong University, Xi’an 710049, China}

\author{Hieu Minh Ngo}%
%\email{hieungo@sogang.ac.kr}
\affiliation{Center for Ultrafast Phase Transformation, Department of Physics, Sogang University, Seoul 04107, Republic of Korea}

\author{Muhammad Mahmood Nawaz}%
%\email{mahmood.cssp@gmail.com}
\affiliation{Center for Ultrafast Phase Transformation, Department of Physics, Sogang University, Seoul 04107, Republic of Korea}

\author{Ana F. Suzana}
%\email{asuzana@anl.gov}
\affiliation{Condensed Matter Physics and Materials Science Department, Brookhaven National Laboratory, Upton, NY 11973, USA}

\author{Longlong Wu}
%\email{lwu@bnl.gov}
\affiliation{Condensed Matter Physics and Materials Science Department, Brookhaven National Laboratory, Upton, NY 11973, USA}

\author{Gareth Nisbet}%
%\email{gareth.nisbet@diamond.ac.uk}
\affiliation{Diamond Light Source, Harwell Science and Innovation Campus, Fermi Ave, Didcot OX11 0DE, UK}

\author{Daniel G. Porter}%
%\email{dan.porter@diamond.ac.uk}
\affiliation{Diamond Light Source, Harwell Science and Innovation Campus, Fermi Ave, Didcot OX11 0DE, UK}

\author{Hyunjung Kim}%
%\email{hkim@sogang.ac.kr}
\affiliation{Center for Ultrafast Phase Transformation, Department of Physics, Sogang University, Seoul 04107, Republic of Korea}

\author{Ian K. Robinson}%
\email{i.robinson@ucl.ac.uk}
\affiliation{Condensed Matter Physics and Materials Science Department, Brookhaven National Laboratory, Upton, NY 11973, USA}
\affiliation{London Centre for Nanotechnology, University College London, London WC1E 6BT, UK}

\date{\today}% It is always \today, today,
             %  but any date may be explicitly specified

\begin{abstract}
Strontium titanate is a classic quantum paraelectric oxide material that has been widely studied in bulk and thin films. It exhibits a well-known cubic-to-tetragonal antiferrodistortive phase transition at 105 K, characterized by the rotation of oxygen octahedra. A possible second phase transition at lower temperature is suppressed by quantum fluctuations, preventing the onset of ferroelectric order. However, recent studies have shown that ferroelectric order can be established at low temperatures by inducing strain and other means. Here, we used \textit{in situ} multi-reflection Bragg coherent x-ray diffraction imaging to measure the strain and rotation tensors for two strontium titanate microcrystals at low temperature. We observe strains induced by dislocations and inclusion-like impurities in the microcrystals. Based on radial magnitude plots, these strains increase in magnitude and spread as the temperature decreases. Pearson's correlation heat maps show a structural transition at 50 K, which could possibly be the formation of a low-temperature ferroelectric phase in the presence of strain. We do not observe any change in local strains associated with the tetragonal phase transition at 105 K.

\begin{comment}
    \begin{description}
    \item[Usage]
    Secondary publications and information retrieval purposes.
    \item[Structure]
    You may use the \texttt{description} environment to structure your abstract;
    use the optional argument of the \verb+\item+ command to give the category of each item. 
    \end{description}
\end{comment}

\end{abstract}

%\keywords{Suggested keywords}%Use showkeys class option if keyword
                              %display desired
\maketitle

%\tableofcontents

% find John Hill's neutron scattering data about STO and properly reference it
\section{\label{sec:Introduction}Introduction}
Strontium titanate (SrTiO$_3$ or STO) is a complex oxide perovskite with mixed ionic and covalent properties, resulting in rich physical phenomena. Its availability in many different forms has made it widely used for various applications, including photocatalysis \cite{Takata2020}, electronics \cite{Cen2009}, and superconductors \cite{Gastiasoro2020}. At room temperature and pressure, it is a cubic perovskite with Sr$^{2+}$ ions at the corners and a Ti$^{4+}$ ion at the center. The Ti$^{4+}$ ion is surrounded by six O$^{2-}$ anions sitting at the center of the cube faces, forming a TiO$_6$ octahedron. 

At low temperatures, STO undergoes notable changes in its structure and properties. At 105 K, it experiences a well-known cubic-to-tetragonal antiferrodistortive (AFD) phase transition \cite{Lytle1964} characterized by the rotation of the TiO$_6$ octahedra \cite{Unoki1967,Fleury1968,Shirane1969}, resulting in a $c/a$ of 1.0009 for the tetragonal unit cell at 10 K \cite{Heidemann1973}. While there is no abrupt phase transition identified, there are several reports of interesting behavior below 50K. The thermal expansion coefficient starts changing around that temperature \cite{Tsunekawa1984}. Below 50 K, it exhibits high dielectric constants that deviate from the classical Curie-Weiss law \cite{Weaver1959}. STO is a classic quantum paraelectric material, meaning it remains paraelectric at low temperatures. The polar modes in such materials soften as the temperature decreases \cite{Shirane1969}, approaching a phase transition but never fully becoming ferroelectric due to suppression from quantum fluctuations \cite{Muller1979}. These quantum fluctuations stabilize the paraelectric phase, preventing the onset of long-range ferroelectric order. 

However, the presence of this quantum suppression supports the description of STO as an incipient ferroelectric. In the ferroelectric phase, the Ti ions shift from their central positions in the oxygen octahedra, breaking the centrosymmetry and creating a dipole moment within the crystal. However, many have suggested that this movement is subdued by the AFD octahedra rotations \cite{Zhong1995,Yamanaka2000}. Nonetheless, ferroelectricity can be introduced by small perturbations such as strain \cite{Haeni2004}, isotope substitution \cite{Itoh1999}, Ca$^{2+}$ substitution \cite{Bednorz1984}, electric fields \cite{Fleury1968,Li2019b}, and laser excitation \cite{Nova2019,Orenstein2024}. Furthermore, the low-temperature phase is positioned close to a quantum critical point, which is related to superconductivity \cite{Enderlein2020}. Understanding the physics behind these phenomena is crucial for the development of next-generation electronic devices.

\begin{figure}
    \centering
    \includegraphics[width=\linewidth]{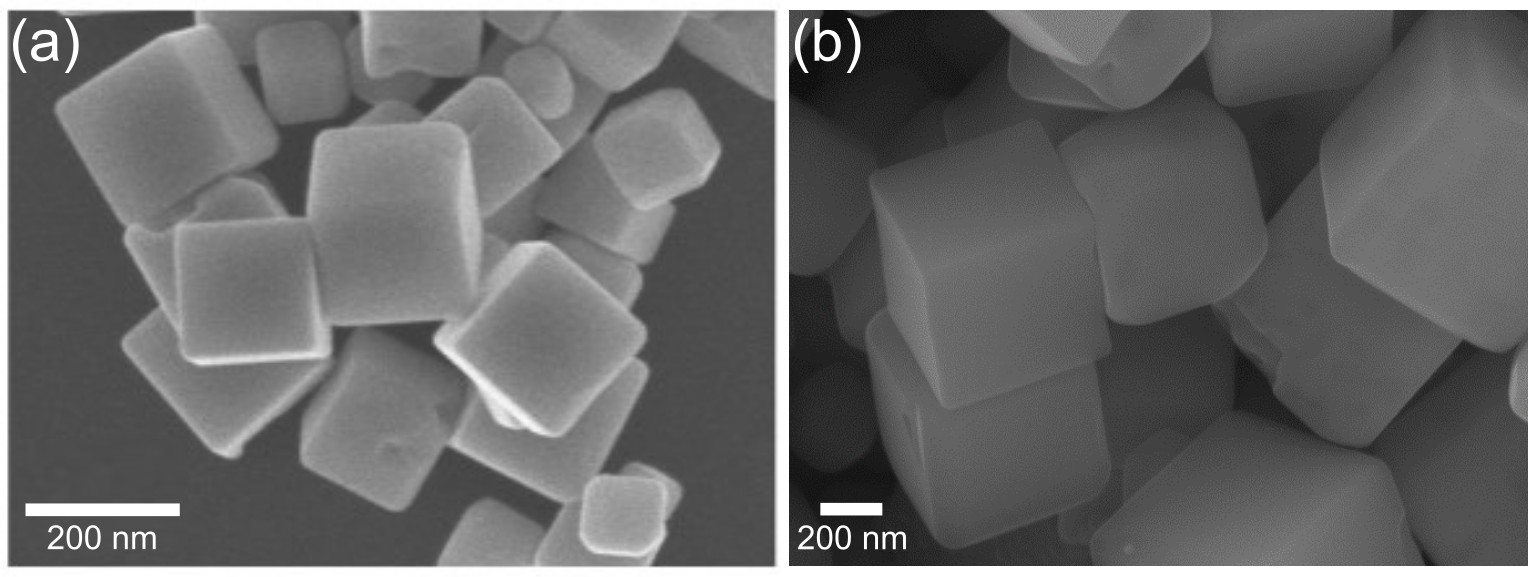}
    \caption{Hydrothermally synthesized STO crystals with cubic morphology. (a) $\sim$300 nm-sized crystals (b) $\sim$600 nm-sized crystals.}
    \label{fig:SEM}
\end{figure}

Many low-temperature studies have been performed on bulk single crystal STO \cite{Weaver1959,Tsunekawa1984,Muller1979,Fleury1968,Lytle1964,Shirane1969,Loetzsch2010,Itoh1999,Li2019b,Nova2019,Enderlein2020} or thin films \cite{He2003,Li2006,Haeni2004}. Investigating the properties of functional materials has recently been focused on the sub-micron level, where surface energy and sample morphology can also dictate the material's properties, emphasizing the need for further exploration. STO microcrystals can be fabricated using solid-state synthesis \cite{Patil2014} and the sol-gel method \cite{Hao2014}, and are widely used in catalysis \cite{Takata2020}. We synthesized STO microcrystals (Fig. \ref{fig:SEM}) using hydrothermal synthesis \cite{Dong2014}, a technique that leverages the solubility differences of metal precursors at elevated temperatures and pressures, leading to the formation of highly crystalline samples that vary in size based on incubation duration and temperature. Of particular interest is the role of strain in dictating the properties of STO microcrystals. For thin film samples, the ferroelectric behavior of STO can be influenced by strain caused by dopants \cite{Muller1991} or epitaxy \cite{Haeni2004}. However, this behavior has not been previously measured in free-standing microcrystals.
 %The coupling between polarization and strain fields is known as flexoelectricity \cite{Zubko2013}, and has been demonstrated in STO \cite{Zubko2007,Gao2018}.  

To probe the strain in microcrystals, we use a non-destructive technique called Bragg coherent x-ray diffraction imaging (BCDI) \cite{Robinson2001, Robinson2009}. BCDI allows for three-dimensional (3D) imaging of isolated micro- and nanocrystalline materials with a spatial resolution of up to 10 nm. It can provide the 3D lattice strain field, projected along the scattering vector, $\mathbf{Q_\mathit{hkl}}$, of a particular $hkl$ crystal reflection, with a resolution on the order of $10^{-4}$ \cite{Carnis2019,Hofmann2020}. BCDI has been recently used for many \textit{in situ} experiments including ZrO grain growth \cite{Dzhigaev2022}, Pt catalysis \cite{Atlan2023a}, Li- and Mn-rich cathode degradation \cite{Liu2022}, and barium titanate phase transformation \cite{Diao2024}.

BCDI involves fully illuminating a crystalline sample inside the coherent volume of an x-ray beam, which is typically no larger than $1\ \mathrm{\mu m}$ in any dimension at third- and fourth-generation synchrotron sources \cite{Robinson2009}. Once the Bragg condition is met for a specific $hkl$ reflection, the diffraction pattern is collected on a pixelated area detector positioned perpendicular to the outgoing wave vector in the Fraunhofer diffraction regime. By rotating the sample about a rocking axis, a 3D coherent x-ray diffraction pattern (CXDP) is collected as the detector moves through the Ewald sphere. If the CXDP is oversampled by at least twice the Nyquist frequency \cite{Sayre1952}, at least 2 pixels per fringe along one dimension, iterative phase retrieval algorithms that apply constraints in real and reciprocal space until convergence is satisfied, can be used to recover the phase \cite{Fienup1982,Fienup1978}. The amplitude and phase in reciprocal space are related to the real-space object via an inverse Fourier transform \cite{Miao2000b,Robinson2001}, followed by a real space transformation from detector-conjugated space to orthogonal laboratory or sample space \cite{Yang2019}. The resulting amplitude, $\mathbf{\rho(r)}$, where $\mathbf{r}$ is the position vector, is proportional to the effective electron density of the crystalline volume associated with the particular crystal reflection. The phase, $\mathbf{\psi(r)}$, corresponds to the projection of the lattice displacement field, $\mathbf{u(r)}$, onto the Bragg vector:

\begin{equation} \label{eq:phase}
    \mathbf{\psi_\mathit{hkl}(r)} = \mathbf{Q_\mathit{hkl}}\cdot\mathbf{u(r)}.
\end{equation}

If at least three linearly independent reflections are collected from a single microcrystal, known as multi-reflection BCDI (MBCDI), one can compute the 3D lattice strain and rotation tensors with respect to an arbitrary reference (Sec. \ref{subsec:Strain_and_rotation_tensors}). Here, we set the center of mass of the reconstructed crystal for each reflection as the zero reference. MBCDI requires knowledge of the crystal orientation, which is typically attained via estimates based on \textit{a priori} crystal geometry \cite{Newton2010}, synchrotron micro-beam Laue diffraction \cite{Hofmann2017b,Pateras2020a}, indexing pole figures corresponding to a known Bragg peak \cite{Richard2018}, and/or laboratory-based electron backscatter diffraction (EBSD) \cite{Yang2022c}. Obtaining multiple reflections for a single crystal can be challenging and labor-intensive, but has been valuable for studying defects and dislocations \cite{Hofmann2020,Phillips2020}, domain structures \cite{Mokhtar2024b} and enabling simultaneous MBCDI phase retrieval procedures to increase reconstruction fidelity \cite{Newton2020,Gao2021,Wilkin2021,Maddali2023}.

Here we explore how defects affect STO microcrystals at low temperature, and how it might differ compared to bulk ($> 1\mathrm{\mu m}$) or thin film STO. We note that some fundamental studies on the low temperature phase in bulk STO have possibly been influenced by sample damage near the surface of the crystal, which contain defects \cite{Wang1998,Shirane1969,Hirota1995,Andrews1986}. With MBCDI, we isolate how these defects affect the low temperature behavior of STO. We study two STO microcrystals, referred to as crystal A and crystal B, at low temperatures. We analyze different structural features in each microcrystal. The first crystal has two dislocations present, which contribute to strain fields that evolve with temperature. The second crystal contains reproducible regions of low electron density, which we refer to as voids, suggesting there is nonuniformity present. Despite these differences, there are temperature trends that remain consistent for both crystals. Although this has recently been accomplished at high temperatures \cite{Chatelier2024}, we use \textit{in situ} MBCDI to determine the residual strain and rotation tensors at cryogenic temperatures, allowing us to capture subtle strain changes ($\sim10^{-4}$) surrounding defects. We use heat maps to examine changes in structure in real and reciprocal space, enabling the detection of different structural modes of STO. Furthermore, the 3D tensor information is analyzed as radial averages, allowing for qualitative monitoring of the strain patterns.

\section{\label{sec:Methods}Methods}

\subsection{\label{subsec:Hydrothermal synthesis}Hydrothermal synthesis}
The synthesis of the STO microcrystals with a cubic shape was inspired by the work of Hao \textit{et al.} \cite{Hao2014}. It first involved the preparation of TiCl$_4$, SrCl$_2$, and LiOH (3M) solutions. Strontium chloride (hexahydrate, 99\%), lithium hydroxide (monohydrate, 98\%), and titanium (IV) chloride (99.9\%) were sourced from Sigma-Aldrich. The TiCl$_4$ solution involved adding 0.26 mL TiCl$_4$ (cooled overnight in a fridge) to a mixture of 10 mL MeOH and 25 mL H$_2$O. The SrCl$_2$ solution was prepared by adding 0.64 g SrCl$_2$ to 10 mL H$_2$O. 3.77 g LiOH was added into 30 mL H$_2$O to form 3M LiOH. Next, we added the TiCl$_4$ and SrCl$_2$ mixture into 3M LiOH to create a white suspension in a Teflon vessel, which was stirred for 30 minutes at room temperature. The sealed Teflon vessel was then positioned in a stainless steel autoclave and treated at 200 °C for 48 h. After, the autoclave was allowed to cool down to room temperature, and the sub-micron crystals were then centrifuged and washed several times with acetone to obtain the powders. The STO particles were then immersed in a 0.035 wt.\% solution of poly(ethyleneimine) in distilled water. This solution was drop-cast onto a silicon substrate, which was then calcined in a furnace at 550 °C for 4 h to complete the sampling process for BCDI measurement. Crystals A and B were from this batch of crystals.

A batch of smaller $\sim$ 300 nm crystals, shown in Fig. \ref{fig:SEM}(a) and \ref{fig:Dipole}, was prepared using a similar procedure \cite{Dong2014} in a different laboratory. The synthesis differed by using an autoclave incubation time of 48 h at 180 °C instead of 48 h at 200 °C.

\subsection{\label{subsec:Data collection}BCDI data collection}
BCDI was performed at beamline I16 at Diamond Light Source (DLS). The silicon substrate was attached to a copper stub using silver paint and placed in a ARS DE-202SK cryocooler. The samples were illuminated using a 9 keV ($\lambda = 0.138\ $ nm) coherent x-ray beam, focused to a size of 50 $\mathrm{\mu m}$ × 200 $\mathrm{\mu m}$ (v × h, full width at half-maximum). Due to cryostat vibrations (0.01 mm, $\sim2$ Hz), we believe having a large beam was beneficial to prevent the microcrystals from moving out of the illuminated spot. The x-ray beam was rastered across the substrate until a suitable specular $\{1\;0\;1\}$ Bragg peak was found. By rotating the sample about the surface normal axis ($\phi$) by 90°, a second $\{1\;0\;1\}$ Bragg peak was obtained, allowing us to determine the orientation matrix to measure additional reflections. 

We accomplished this for two crystals, which were selected based on having strong Bragg peak signals. For crystal A, we measured the $10\bar{1}$, $110$, $101$, $1\bar{1}0$, and $111$ Bragg peaks at 10, 20, 30, 40, 50, 100, 150, 200, 250, and 300 K. However, the reconstruction for the $111$ Bragg peak did not converge as well and was therefore excluded from the analysis, except for the determination of the Burgers vector (Table \ref{table:Q_dot_b}). For crystal B, the $111$ and $011$ Bragg peaks were measured at 180, 160, 140, 100, 80, 50, 40, 30, 20 and 10 K, and the $101$ Bragg peak was measured at 180, 160, 140, 100, 80, and 50 K. The temperature ramp rate was 5 K/min, and once a specific temperature was reached, 5 minutes elapsed before starting the BCDI measurements. The sample temperature was measured using a calibrated diode attached to the copper stub. The sample was kept in the beam as the cryostat thermally contracted by correcting the height using a previously measured calibration curve.

Coherent x-ray diffraction patterns (CXDPs) were collected on a 512 × 512 pixel Quad-Merlin detector with a pixel size of $55 \ \mathrm{\mu m} \times 55 \ \mathrm{\mu m}$, positioned at 1.31 m from the sample to ensure oversampling. CXDPs were recorded by rotating the crystal through an angular range of 0.5° - 0.6° and recording an image every 0.005° with a 1 s exposure time. To increase the signal-to-noise ratio, five repeated scans for each of the reflections at every temperature were measured.

\subsection{\label{subsec:Data processing}Data processing}
The CXDP was cropped to a size of $166 \times 166$ pixels in the detector plane to avoid detector module gaps during data processing. Each set of five repeated scans was aligned using a 3D version of a 2D subpixel translation procedure proposed by Guizar-Sicairos \textit{et al.} \cite{Guizar-Sicairos2008}. After alignment, the Pearson correlation of the set of repeated CXDPs was greater than 0.99, demonstrating that the crystal and environmental setup remained remarkably stable in the cryostat for most of the measurements. The scans were summed, and the background was determined as the mean value of a fitted histogram of the summed data. Before fitting the histogram, values below 1 and those greater than 95\% of the maximum intensity were removed. The background of the five summed scans was subtracted from the CXDP and also used as the minimum data threshold to recover the complex electron density in the phase retrieval algorithm. Further details can be found in Appendix \ref{appendix:phase_retrieval}.

\subsection{\label{subsec:Strain_and_rotation_tensors}Residual strain and rotation tensor computation} 
Based on infinitesimal strain theory, the 3D lattice strain tensor, $\mathbf{\epsilon}(\mathbf{r})$, and rotation tensor, $\mathbf{\omega}(\mathbf{r})$, with respect to an arbitrary reference, are given by \cite{Constantinescu2008}:

\begin{align}
    \begin{split} \label{eq:lattice_strain_tensor}
        \mathbf{\epsilon}(\mathbf{r}) &= \frac{1}{2}\left\{\nabla \mathbf{u}(\mathbf{r})+[\nabla\mathbf{u}(\mathbf{r})]^\top\right\}\mathrm{,\ and}\\
    \end{split}\\\nonumber\\
    \begin{split} \label{eq:rotation_tensor}
        \mathbf{\omega}(\mathbf{r}) &= \frac{1}{2}\left\{\nabla \mathbf{u}(\mathbf{r})-[\nabla\mathbf{u}(\mathbf{r})]^\top\right\}.\\
    \end{split}
\end{align}

Here we found $\nabla \mathbf{u}(\mathbf{r})$ directly for the computation of $\mathbf{\epsilon}(\mathbf{r})$ and $\mathbf{\omega}(\mathbf{r})$ in Eqs. \ref{eq:lattice_strain_tensor} and \ref{eq:rotation_tensor}, respectively. Using the modified approach by Hofmann \textit{et al.} \cite{Hofmann2020}, we minimized the squared error between phase gradients,

\begin{equation} \label{eq:least-squares_gradient}
    E(\mathbf{r})_j = \sum_{hkl,j} \left[\mathbf{Q}_{hkl}\cdot\frac{\partial \mathbf{u}(\mathbf{r})}{\partial j}-\frac{\partial \psi_{hkl}(\mathbf{r})}{\partial j}\right]^2,
\end{equation}

where $j$ corresponds to the spatial $x$, $y$, or $z$ sample coordinates. The strain and rotation tensor components, $\mathbf{\epsilon}$ and $\mathbf{\omega}$ in this paper were computed relative to the sample coordinate systems in crystal A (Fig. \ref{fig:Tensor_slices_A}) and crystal B (Fig. \ref{fig:Tensor_slices_B}) respectively. The phase gradients were computed using the derivative of the complex exponential of the phase \cite{Yang2022c} to account for phase jumps resulting from dislocations with characteristic phase vortices \cite{Clark2015}.

\section{\label{sec:Results}Results}
\subsection{\label{subsec:Crystal_A}Crystal A with dislocations}

\begin{figure}
    \centering
    \includegraphics[width=\linewidth]{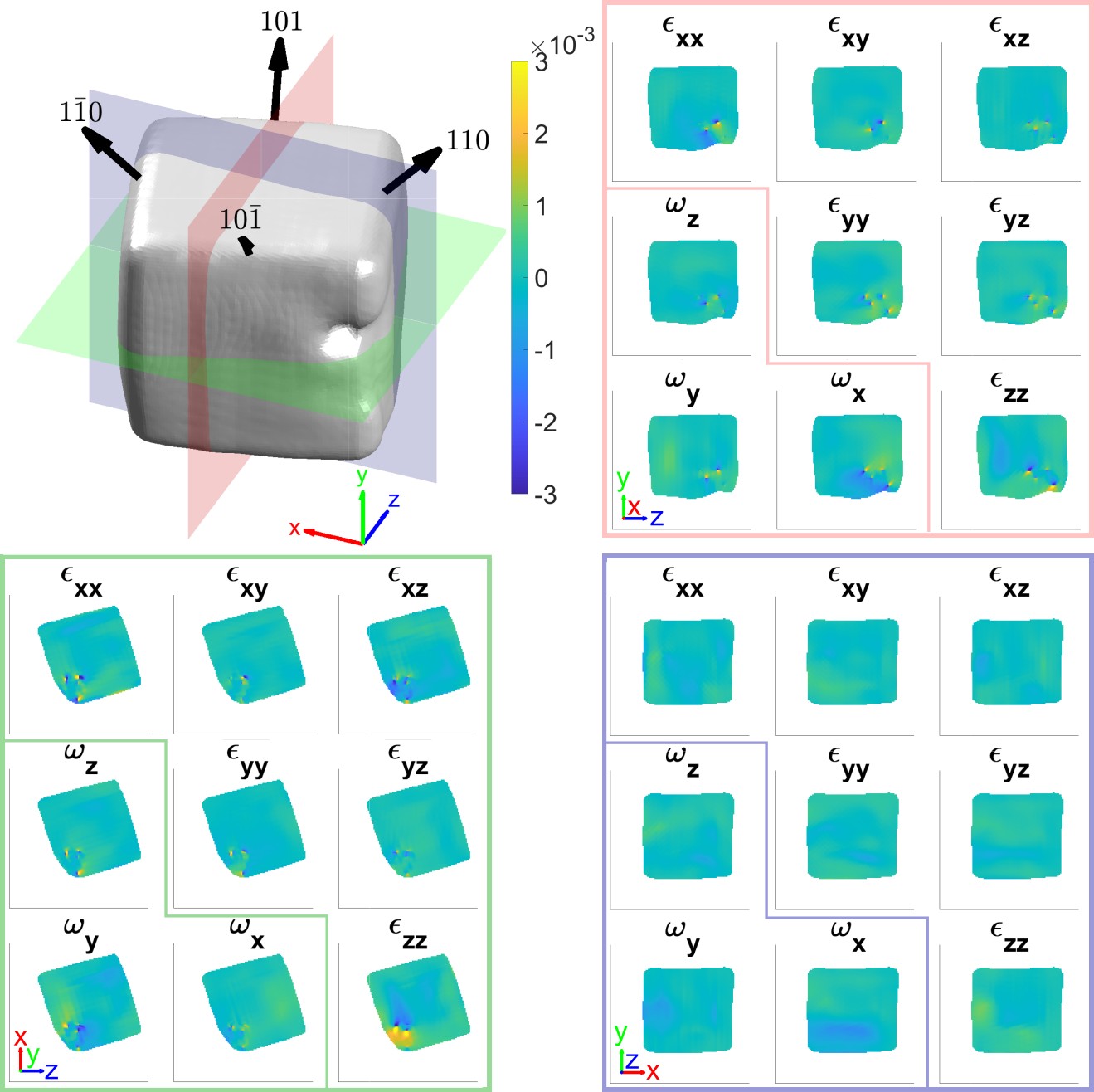}
    \caption{The average morphology of the $10\bar{1}$, $110$, $101$, and $1\bar{1}0$ reflections for STO crystal A at 300 K, based on a normalized amplitude threshold of 0.20. Slices through the residual strain and rotation tensor components using the average morphology are shown for the planes indicated at $x = 2.5$ nm (red), $y = 2.5$ nm (green), and $z = 2.5$ nm (blue) from the center of mass of the microcrystal. The coordinate axes arrows have a length of 100 nm. Supplemental Material videos 1–3 \cite{supp} show the residual strain and rotation tensor components throughout the volume along the $x$, $y$, and $z$ axes, respectively.}
    \label{fig:Tensor_slices_A}
\end{figure}

The tensor components and average morphology for crystal A are shown in Fig. \ref{fig:Tensor_slices_A}. This crystal is 400 nm in length and exhibited lattice strain patterns in the \textit{y-z} and \textit{x-z} planes, indicating the presence of dislocations \cite{Hofmann2020}. This observation was further confirmed by the phase vortices in each measured reflection, shown in Appendix \ref{appendix:Individual_reconstructions}.

\subsubsection{\label{subsubsec:Dislocations}Crystal A dislocation identification}

\begin{figure}
    \centering
    \includegraphics[width=\linewidth]{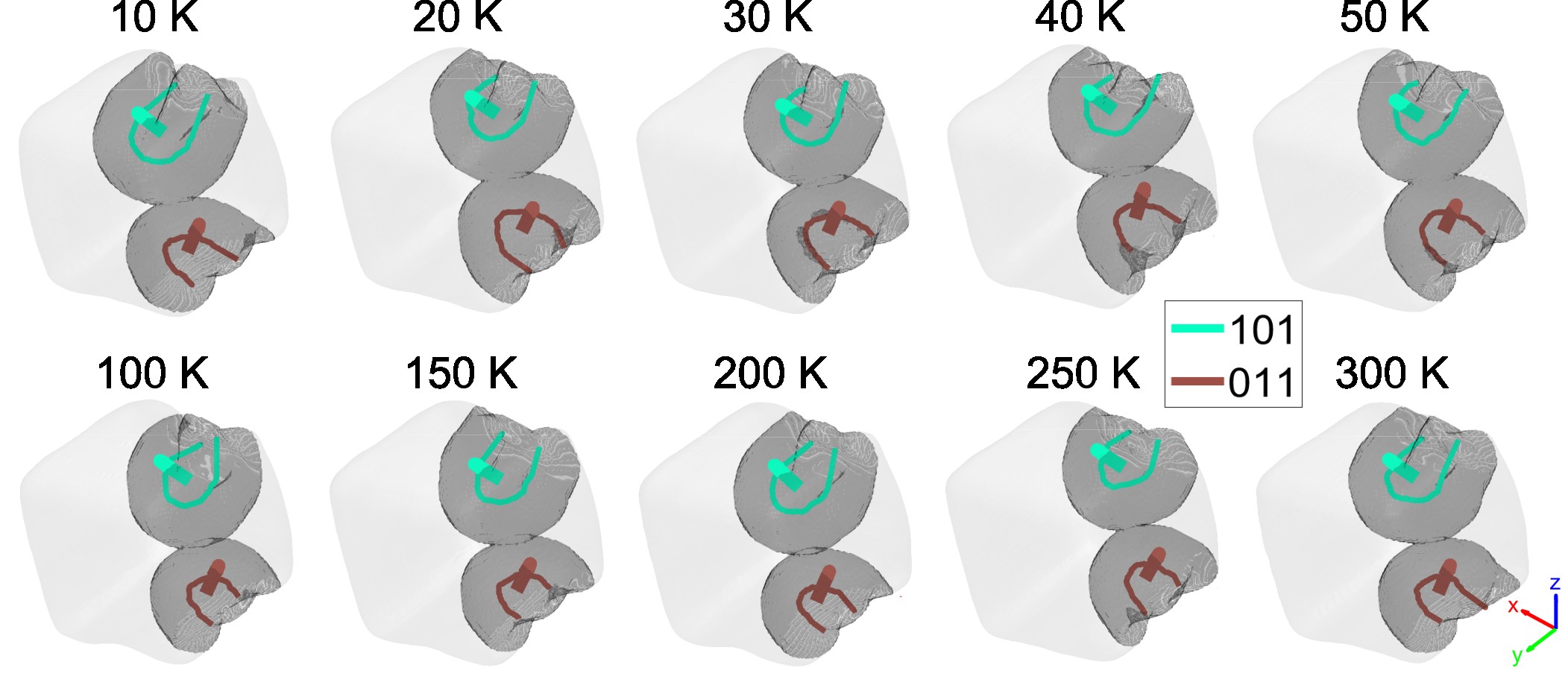}
    \caption{The average positions of the dislocation lines. From the dislocation analysis (Appendix \ref{appendix:Dislocation_analysis}), the top dislocation (green) has a Burgers vector of $\mathbf{b} = a[1\;0\;1]$, and the bottom dislocation (brown) has a Burgers vector of $\mathbf{b} = a[0\;1\;1]$. The colored arrows indicate the normal to the dislocation plane. The dark region corresponds to the dislocation region or mask used for radial analysis in Sec. \ref{subsubsec:Radial_tensor_A}. The coordinate axes arrows have a length of 100 nm. The dislocations are discussed in more detail in Appendix \ref{appendix:Dislocation_analysis}.}.
    \label{fig:Dislocation_lines_avg}
\end{figure}

To identify the positions of the dislocations at each temperature, we averaged the positions of the dislocations in each reflection as discussed in Appendix \ref{appendix:Dislocation_analysis}. If averaging the dislocation lines did not result in dislocations that terminated at the surface of the crystal, they were extended to the surface. The resulting average dislocation lines are shown in Fig. \ref{fig:Dislocation_lines_avg} and lie on the $(1\;0\;1)$ and $(0\;1\;1)$ planes for the top and bottom dislocations, respectively. Based on Table \ref{table:Q_dot_b} in Appendix \ref{appendix:Dislocation_analysis}, the Burgers vectors were $\mathbf{b} = a[1\; 0\; 1]$ and $\mathbf{b} = a[0\; 1\; 1]$ for the top and bottom dislocations, respectively, indicating that they are prismatic dislocation loops. These dislocations were likely formed during the hydrothermal synthesis process.

%, suggesting that they were screw dislocations. It is not surprising to find this type of dislocation in STO, as the one of the most common slip systems is of type $<1\;1\;0>\{\bar{1}\;1\;0\}$. 

We created a ``dislocation mask'', shown as a dark region surrounding each dislocation in Fig. \ref{fig:Dislocation_lines_avg}. This region was created by masking a 90 nm radius along each average dislocation line and covered roughly 25\% of the average crystal morphology volume. Its purpose was to exclude most of the lattice strain contributions from the dislocation core, as discussed in Sec. \ref{subsubsec:Radial_tensor_A}.
%0.27, 0.26, 0.25, 0.25, 0.19, 0.24, 0.07, 0.25, 0.22, 0.18

\subsubsection{\label{subsubsec:XC_A}Crystal A Pearson correlation coefficient}
\begin{figure}
    \centering
    \includegraphics[width=\linewidth]{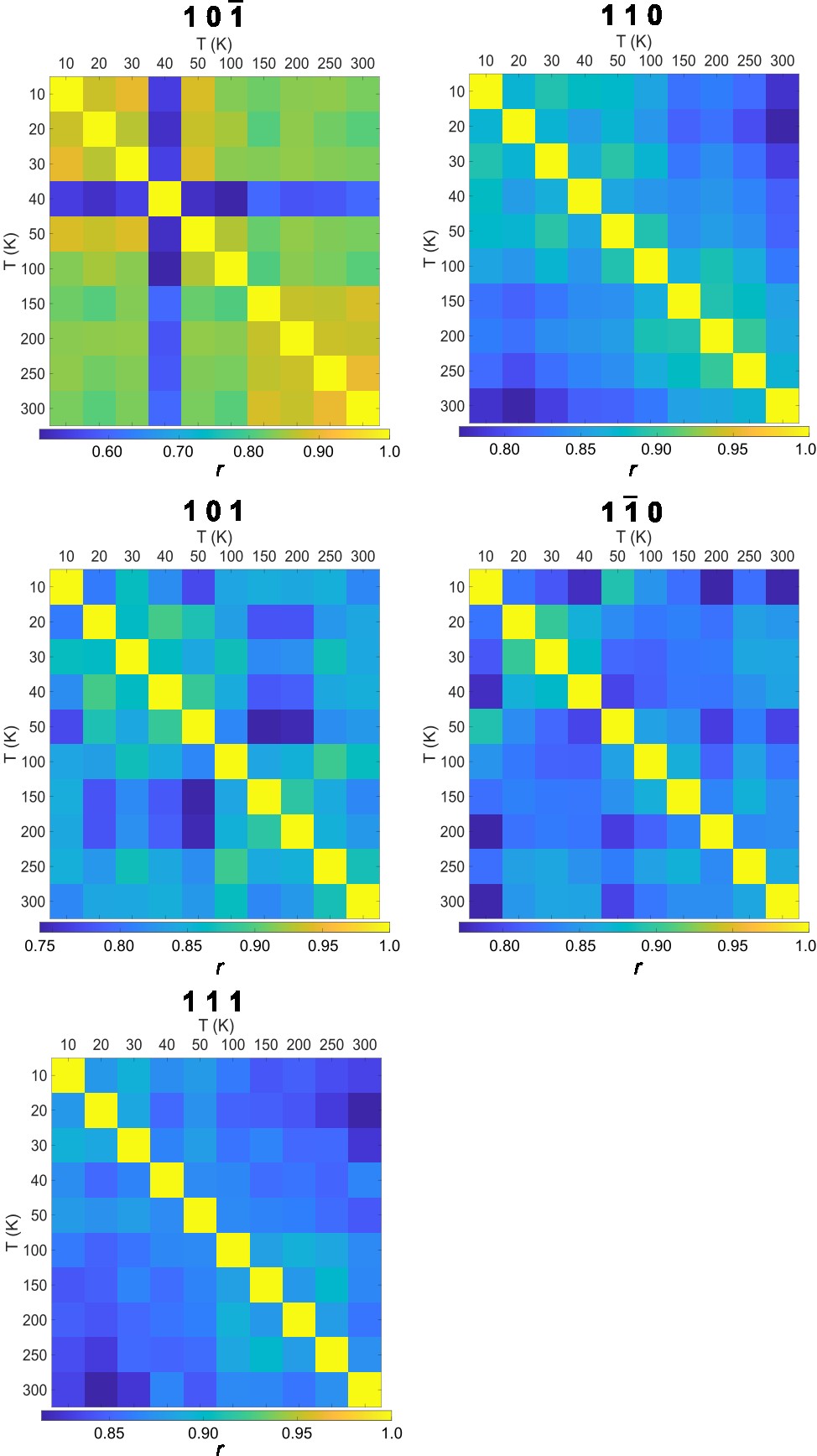}
    \caption{Pearson's $r$ correlation coefficient (Eq. \ref{eq:XC}) heat maps of the Bragg peak intensity distribution (binarized) for different temperatures and reflections. Note that the crystal was misaligned during the measurement of the $10\bar{1}$ CXDP at 40 K.}
    \label{fig:CXDP_XC_A}
\end{figure}

To monitor changes in the crystal as a function of temperature, we employed the Pearson correlation coefficient, $r$. This method has been used in BCDI \cite{Yau2017a, Ulvestad2015c, Kim2018d, Yang2021, Choi2020} to infer structural transitions, and is the basis underlying the x-ray Photon Correlation Spectroscopy (XPCS) method for studying dynamics of fluctuations \cite{Dierker1995}. The $r$ value between temperatures was computed using Eq. \ref{eq:XC}:

\begin{equation}\label{eq:XC}
     r(x,y) = \frac{\sum\limits_{n}(x_n-\bar{x})(y_n-\bar{y})}{\sqrt{\sum\limits_{n}(x_n-\bar{x})^2}\sqrt{\sum\limits_{n}(y_n-\bar{y})^2}}
\end{equation}

where $x$ and $y$ were the temperatures being compared, $x_n$ and $y_n$ were the values for a single voxel, and $\bar{x}$ and $\bar{y}$ were the means of each array. Prior to computing Pearson's coefficient for the CXDPs, we processed them based on the procedure listed in Sec. \ref{subsec:Data processing}. Afterward, the CXDPs were binarized by setting values greater than zero to one, thereby ignoring the intensity values for Pearson correlation comparison. Fig. \ref{fig:CXDP_XC_A} shows the trends associated with each reflection and temperature for crystal A.

\begin{figure}
    \centering
    \includegraphics[width=\linewidth]{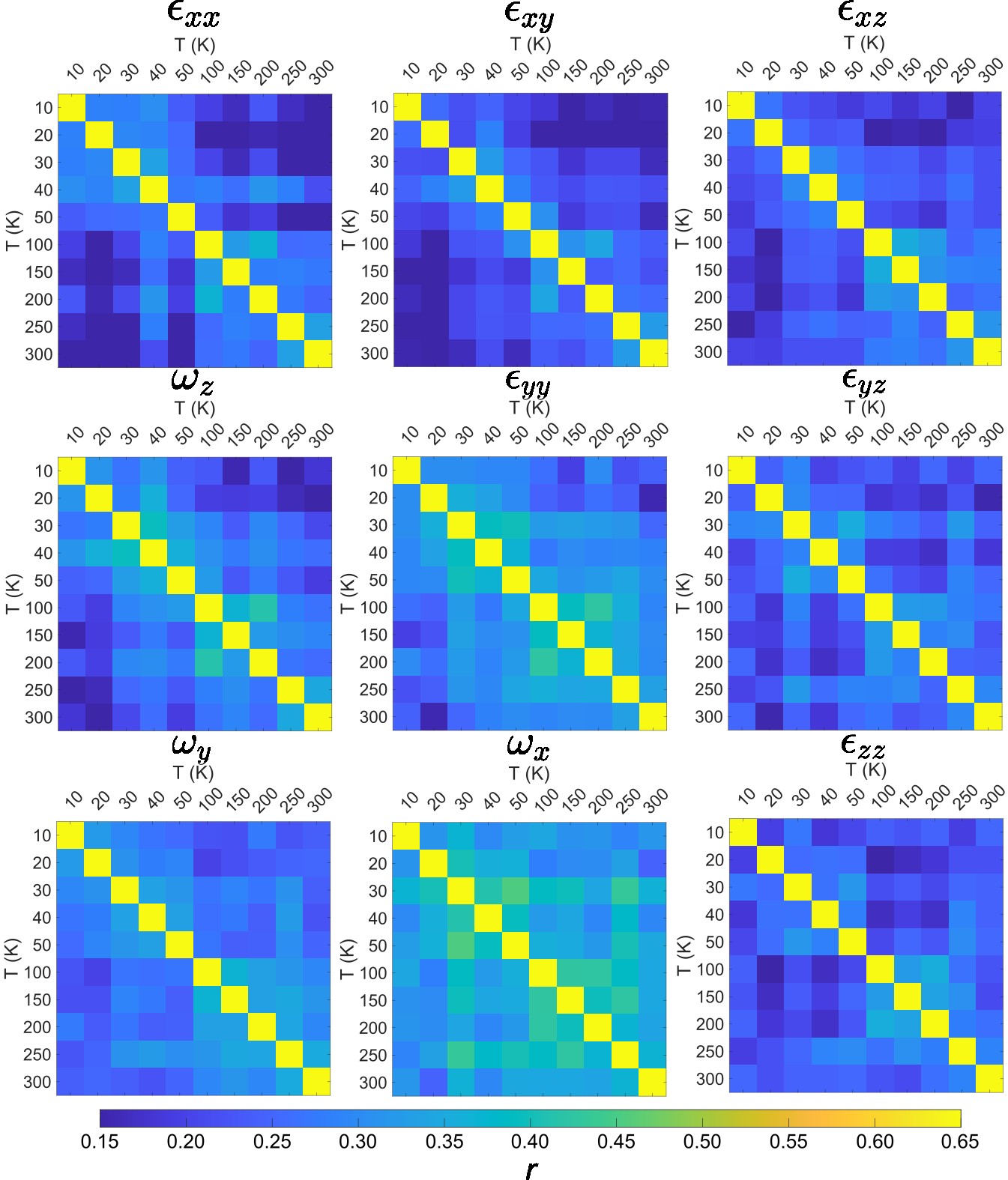}
    \caption{Pearson's $r$ correlation coefficient (Eq. \ref{eq:XC}) heat maps of the tensor components for different temperatures.}
    \label{fig:Tensor_XC_A}
\end{figure}

Similarly, the Pearson correlation was applied to the tensor component values, as shown in Fig. \ref{fig:Tensor_XC_A}. 

\subsubsection{\label{subsubsec:Radial_tensor_A}Crystal A radial strain and rotation tensor magnitudes}
\begin{figure*}
    \centering
    \includegraphics[width=\linewidth]{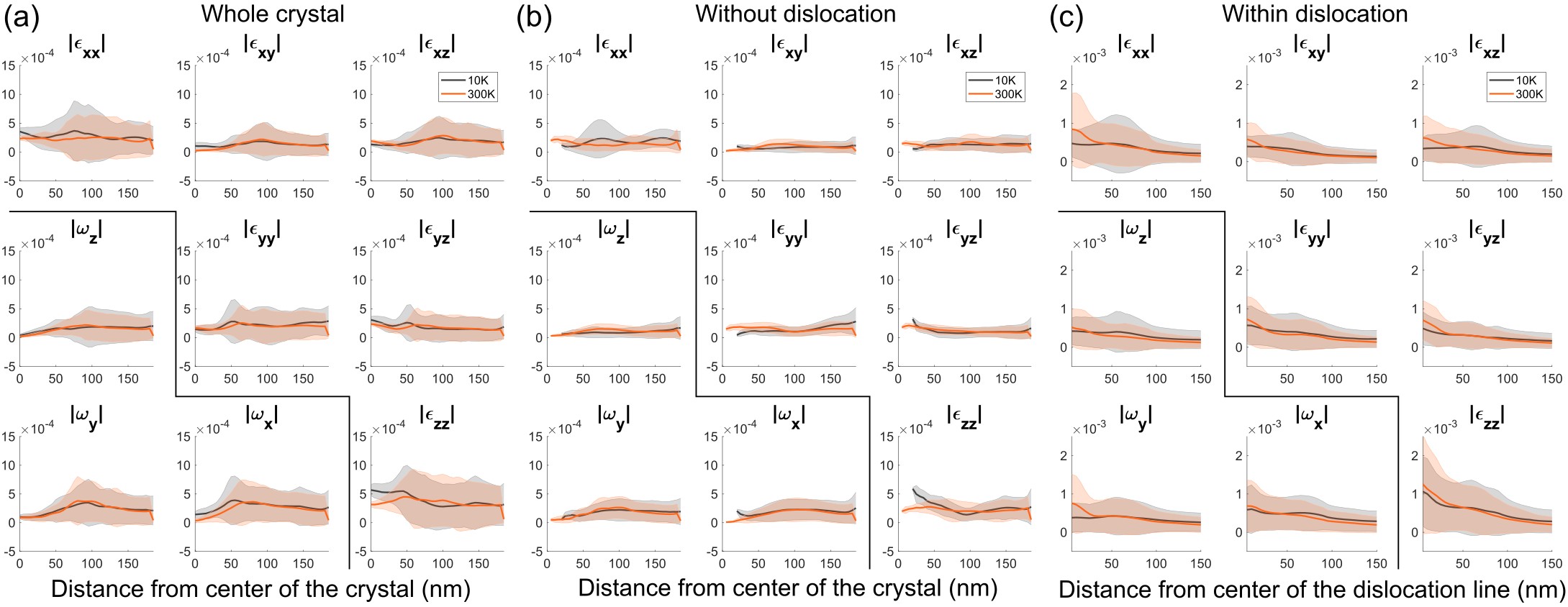}
    \caption{Histogram of the radial distribution of the tensor magnitudes for crystal A at 10 K (black) and 300 K (red). Radial averages are computed for (a) the entire crystal, (b) excluding the dislocation region shown in Fig. \ref{fig:Dislocation_lines_avg}, and (c) within the dislocation region. The shaded areas in the plots correspond to the standard deviation of the mean at each radial position from the center of the crystal for (a) and (b) and from the center of the dislocation line for (c). Note that for (b), up to the first 20 nm from the center is within the dislocation mask.}
    \label{fig:Radial_strain_A}
\end{figure*}

Using the extreme ends of the temperature measurements, 10 and 300 K, we compared the differences in the tensor components at these two temperature states. To quantify the strain distribution within the crystal at different temperatures, the magnitude of the tensor component was averaged over successively larger shells from the center of the crystal \cite{Clark2015,Yang2022a}. For each radial position, the average magnitude of the tensor components, along with the standard deviation of the magnitudes, is plotted in Fig. \ref{fig:Radial_strain_A}(a). In Figs. \ref{fig:Radial_strain_A}(b) and \ref{fig:Radial_strain_A}(c), this distribution was subdivided into regions of strain located close to the dislocations and far away, by use of the dislocation mask shown in Fig. \ref{fig:Dislocation_lines_avg}. %Separating the contributions of the dislocations from the remainder of the material might highlight the signatures of the STO phase transition.

All the strain tensor components in Fig. \ref{fig:Radial_strain_A} are stronger close to the dislocations. The fact that the dislocations occupied the outermost regions of the crystal explains the shape of the overall strain distribution in Fig. \ref{fig:Radial_strain_A}(a): the drop in average strain below 50 nm radius is linked to the relative absence of dislocations there. Up to 50 nm from the center of the crystal, we observe relatively low strain magnitudes and low heterogeneity, as the dislocation strain fields were too weak. From 50 nm to the surface of the crystal, the average magnitudes and heterogeneity increase as the dislocation strain fields became more pronounced in this region. Within the dislocation, the tensor component magnitudes decrease away from the dislocation line, shown in in Fig. \ref{fig:Radial_strain_A}(c).

The magnitude is slightly higher at 10 K compared to 300 K, especially close to the center of the crystal where no dislocations are present. To further analyze this, we computed the relative volume change, or volumetric strain, $\epsilon_{\mathrm{vol}}$, defined as $\epsilon_{\mathrm{vol}} = \epsilon_{xx} + \epsilon_{yy} + \epsilon_{zz}$.

\begin{figure}
    \centering
    \includegraphics[width=\linewidth]{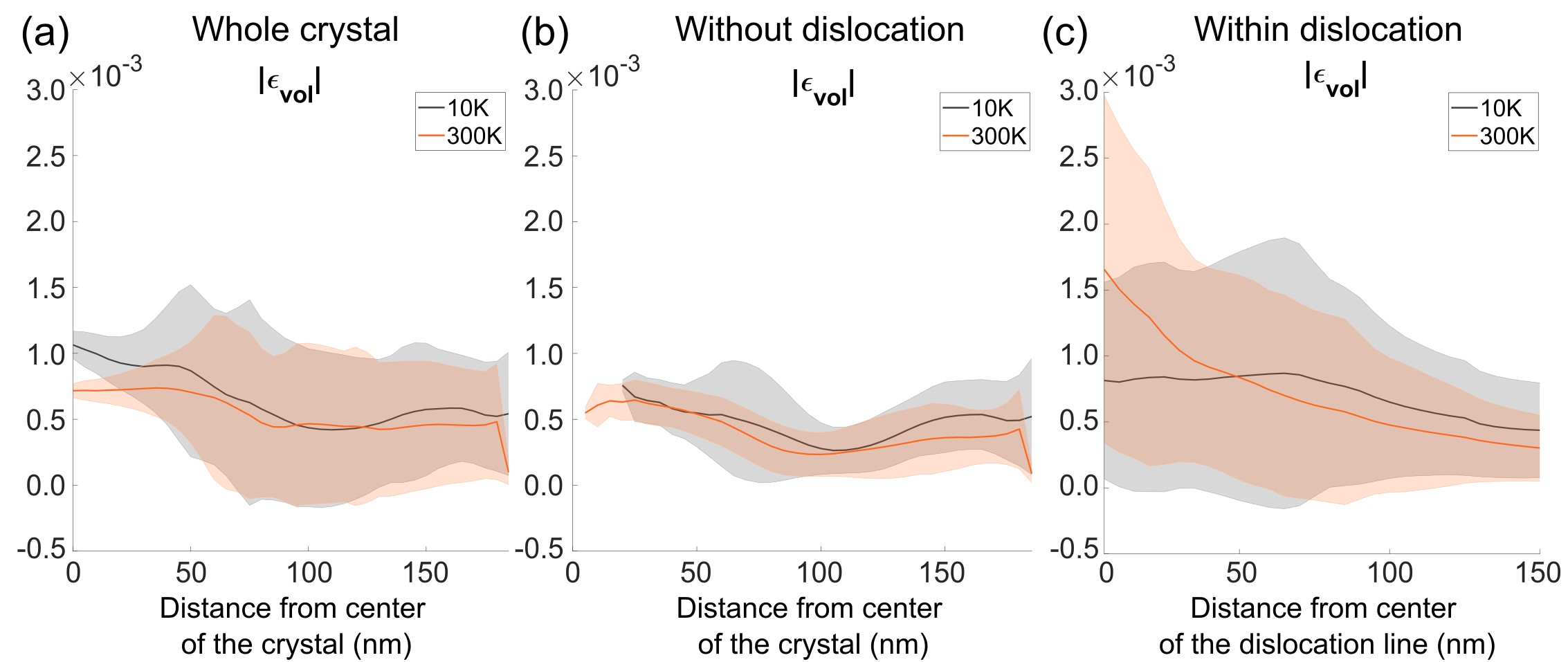}
    \caption{Histogram of the radial distribution of the volumetric strain magnitudes, $\epsilon_{\mathrm{vol}}$, magnitudes for crystal A at 10 K (black) and 300 K (red). Radial averages are computed for (a) the entire crystal, (b) excluding the dislocation region shown in Fig. \ref{fig:Dislocation_lines_avg}, and (c) within the dislocation region. The shaded areas in the plots correspond to the standard deviation of the mean at each radial position from the center of the crystal for (a) and (b) and from the center of the dislocation line for (c). Note that for (b), up to the first 20 nm from the center is within the dislocation mask.}
    \label{fig:Radial_strain_vol_A}
\end{figure}

Fig. \ref{fig:Radial_strain_vol_A} shows similar trends to those in Fig. \ref{fig:Radial_strain_A}, noting that there is an increased volumetric strain present at low temperature, as shown in Fig. \ref{fig:Radial_strain_vol_A}(c). %This could be due to the antiferrodistortive phase transition in STO, caused by the rotation of the TiO$_{6}$ octahedra.

\subsection{\label{subsec:Crystal_B}Crystal B with voids}
To confirm the results in Sec. \ref{subsec:Crystal_A}, a second crystal from the same synthesis batch was measured in a separate experiment at different temperatures and for different reflections. The tensor components and average morphology for crystal B are shown in Fig. \ref{fig:Tensor_slices_B}. The individual reflections used for this computation are discussed in Appendix \ref{appendix:Individual_reconstructions}. This crystal is 600 nm in length and, unlike crystal A, showed no signs of dislocations. Rather, crystal B contained spherical-shaped regions where the effective electron density falls well below that of the rest of the crystal, which we consider to be ``voids''. These voids, highlighted as colored circles in Fig. \ref{fig:Tensor_slices_B} in the \textit{y-z} plane, appeared in slightly different locations for different reflections, as indicated by the colors. %However, these voids are reproducible 

\begin{figure}
    \centering
    \includegraphics[width=\linewidth]{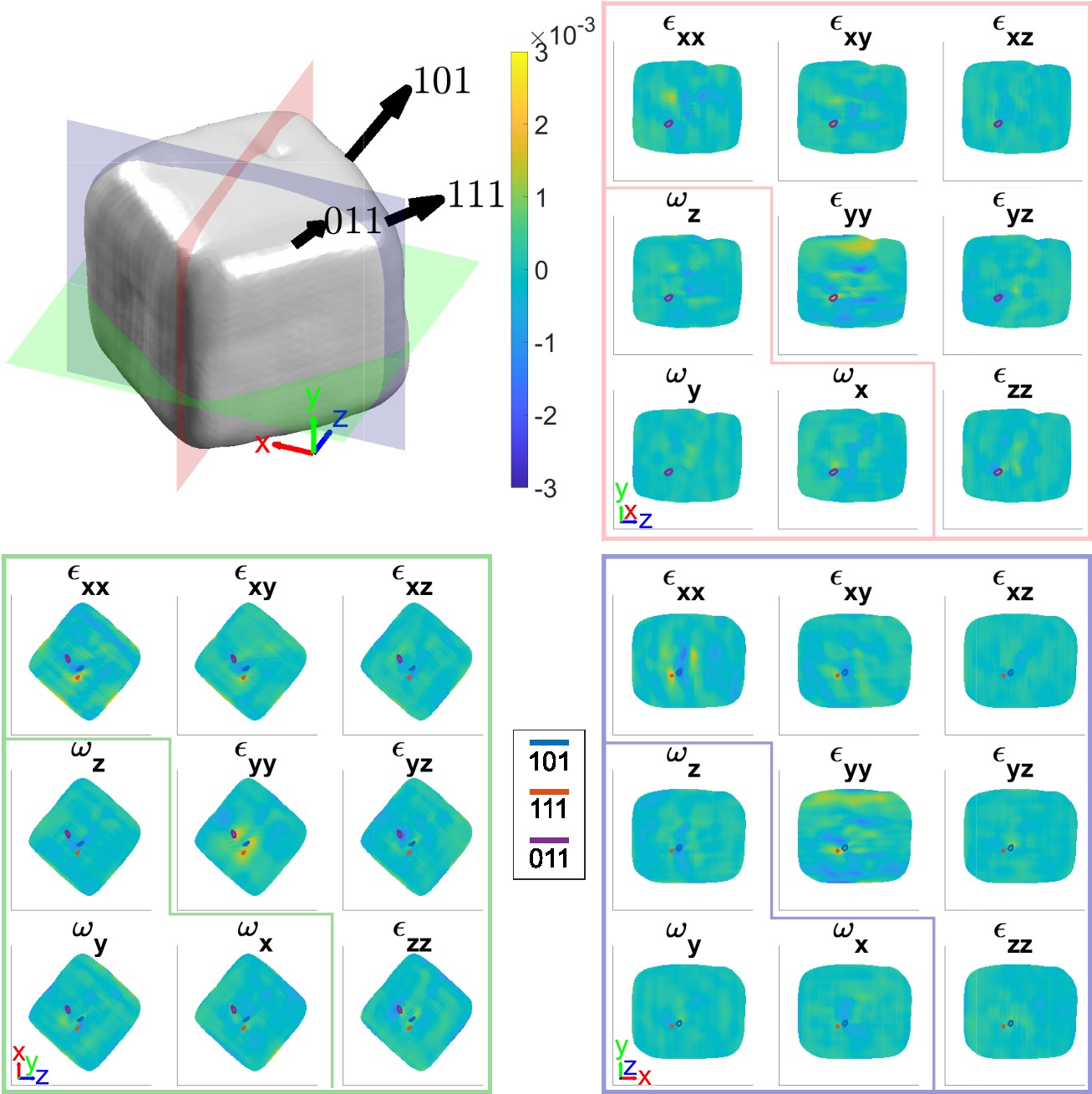}
    \caption{The average morphology of the $101$, $111$, and $011$ reconstructions for STO crystal B at 180 K, based on a normalized amplitude threshold of 0.20. Slices through the residual strain and rotation tensor components using the average morphology are shown for the planes indicated at $x = 2.5$ nm (red), $y = -92.5$ nm (green), and $z = -42.5$ nm (blue) from the center of mass of the microcrystal. The legend indicates the regions of amplitude below the threshold for each reflection, which is shown in the slices along the x direction. The coordinate axes arrows have a length of 100 nm. Void outlines colored for for each reflection are shown in the tensor slices. Supplemental Material videos 4–6 \cite{supp} show the residual strain and rotation tensor components throughout the volume along the $x$, $y$, and $z$ axes, respectively.}
    \label{fig:Tensor_slices_B}
\end{figure}

\begin{figure}
    \centering
    \includegraphics[width=\linewidth]{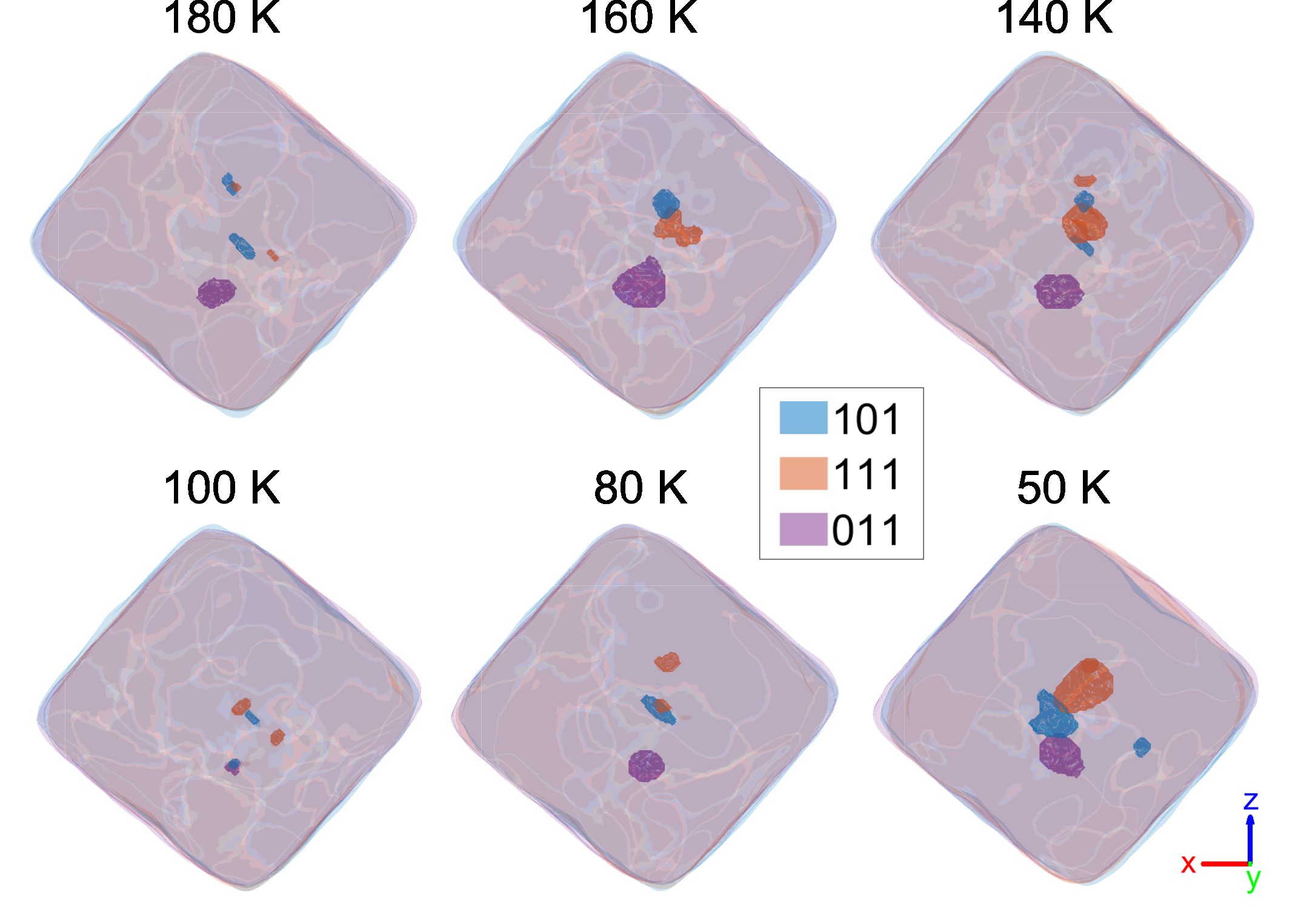}
    \caption{Crystal B translucent amplitude overlap, illustrating the volumes of missing electron density for different reflections. The normalized amplitude threshold is 0.20, and the coordinate axes have a length of 100 nm.}
    \label{fig:Void_overlap}
\end{figure}

Fig. \ref{fig:Void_overlap} shows a translucent overlap of the normalized amplitude for the three measured reflections. Despite strong overlap along the surfaces of each reconstruction, there is poor overlap for the voids at each temperature. However, the voids for each reflection have relatively consistent positions across different temperatures, though their sizes are slightly different. The size differences were due to minor variations in the reconstructed amplitude surrounding the voids. The voids only represent the amplitude region that falls below the threshold of 0.20, but the volume of lower amplitude likely is larger than the void shown in each reflection and can overlap with the other volumes of lower amplitude in the other reflections (see Fig. \ref{fig:Individual_reconstructions_B_amp} in Appendix \ref{appendix:Individual_reconstructions}). The lack of electron density information in the reconstructions implies that the void regions were either not diffracting under the given Bragg conditions or the information was not captured by the detector. Since the voids appear uniquely in certain reflections, one possibility is that the center of crystal B is composed of multiple smaller, crystalline impurities, forming the seed during hydrothermal synthesis. These smaller impurities would diffract under certain Bragg conditions, similar to twin domains within a crystal \cite{Ulvestad2015b}.

\subsubsection{\label{subsubsec:SrCO3}Crystal B SrCO$_{3}$ impurities}
\begin{figure}
    \centering
    \includegraphics[width=\linewidth]{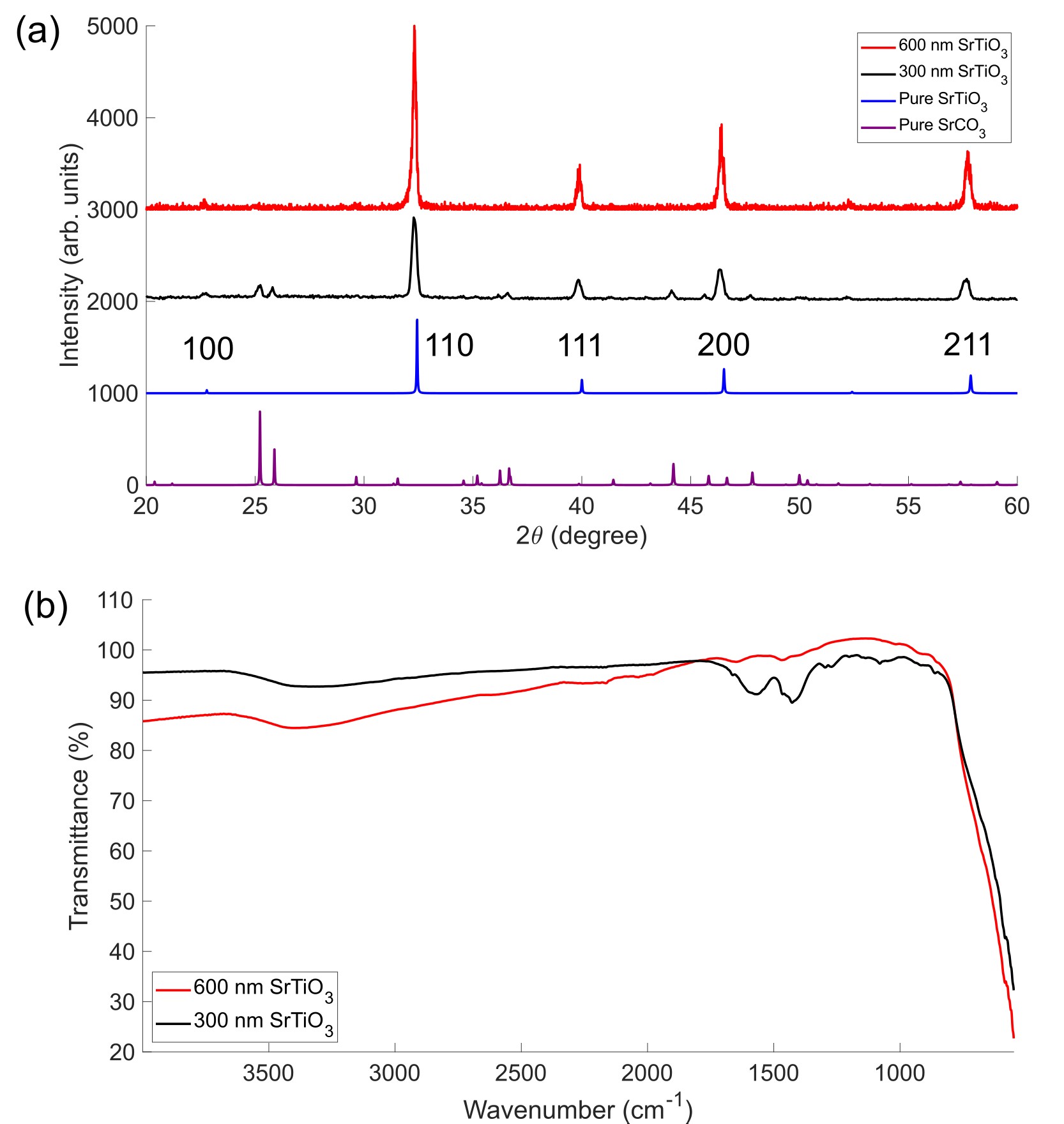}
    \caption{Chemical characterization of the hydrothermally synthesized STO samples. (a) Powder XRD spectra of crystals of different sizes compared to references. The average crystal size based on the Scherrer formula is 614 and 349 nm for each batch. Peaks corresponding to SrCO${_3}$ impurities are observed. (b) FTIR spectra of crystals of different sizes. SrCO${_3}$ impurities appear as absorption bands, most noticeable around 1440 cm$^{-1}$.}
    \label{fig:XRD_FTIR}
\end{figure}

To determine whether the void regions are twins or impurities, we analyzed the samples using x-ray diffraction (XRD) and Fourier-transform infrared spectroscopy (FTIR). Both methods showed that SrCO$_{3}$ was present in the samples. Powder XRD (Fig.\ \ref{fig:XRD_FTIR}(a)) indicated the presence of SrCO$_{3}$ in the STO spectrum, particularly noticeable in the 300 nm microcrystals. The FTIR spectrum revealed characteristic carbonate absorption bands, notably around 867, 1067, and 1440 cm$^{-1}$, which are due to CO$_3^{2-}$ groups \cite{Bacha2011}.

\begin{figure}
    \centering
    \includegraphics[width=\linewidth]{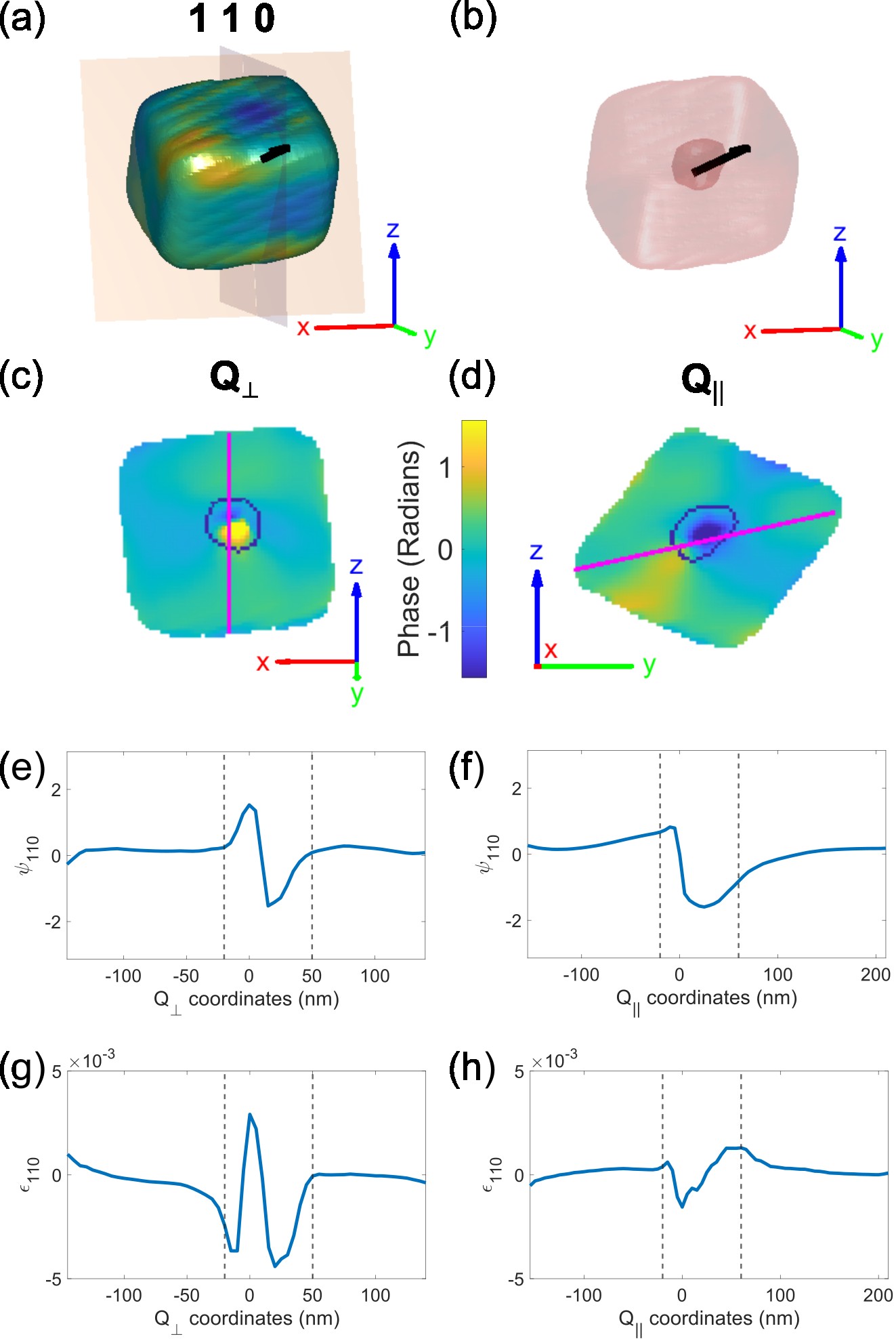}
    \caption{Characterization of a single region with lower effective electron density in a 200-nm STO crystal. (a) An isosurface rendering colored by phase with slices perpendicular and parallel to the scattering vector, $\mathbf{Q_\mathit{hkl}}$ (black arrow). (b) A translucent isosurface from (a) showing the void with an amplitude threshold of 0.20. (c) A slice through the phase of the STO crystal, as shown in (a), perpendicular to $\mathbf{Q_\mathit{hkl}}$ (pointing out of the page). (d) A slice through the phase of the STO crystal, as shown in (a), parallel to $\mathbf{Q_\mathit{hkl}}$. The black ring corresponds to the void outline in (c) and (d). (e) A line profile, indicated by the magenta line in (c), of the phase along $[1\;1\;0]$ of the STO crystal perpendicular to $\mathbf{Q_\mathit{hkl}}$. (f) A line profile, indicated by the magenta line in (d), of the phase along $[1\;1\;0]$ of the STO crystal parallel to $\mathbf{Q_\mathit{hkl}}$. (g) The $[1\;1\;0]$ strain corresponding to the phase in (e). (h) The $[1\;1\;0]$ strain corresponding to the phase in (f). The magnitude of the coordinate axes corresponds to a length of 100 nm for (a)-(d), and the phase color bar ranges from $-\frac{\pi}{2}$ to $\frac{\pi}{2}$.}
    \label{fig:Dipole}
\end{figure}

The introduction of CO$_3^{2-}$ in hydrothermally synthesized STO is believed to originate from atmospheric CO$_2$ reacting with the bases involved in the process \cite{Bacha2011}. The central location of these SrCO$_{3}$ impurities in Fig. \ref{fig:Void_overlap} suggests that they precipitated first and became part of the nucleation seed from which crystal B grew during the hydrothermal synthesis. This SrCO$_{3}$ seed appears localized but complex in shape. It is no larger than 200 nm in size, though it could be composed of multiple small SrCO$_{3}$ impurities at the center of crystal B. Among a small number of examples studied, the presence of voids was not consistent. They were not present at all in crystal A, and tended to be more common and larger in the 300 nm batch, from which one example is shown in Fig. \ref{fig:Dipole}. Here, there is a clear phase dipole parallel to the $\mathbf{Q_\mathit{hkl}}$ direction, indicating there was a pressure difference between the void and the STO crystal, which could arise from differential expansion between different materials. When combined with other reflections to form the strain tensor, this associated phase lead to strain patterns, such as those already seen for crystal B in Fig. \ref{fig:Tensor_slices_B}.

\subsubsection{\label{subsubsec:XC_B}Crystal B Pearson correlation coefficient}
\begin{figure}
    \centering
    \includegraphics[width=\linewidth]{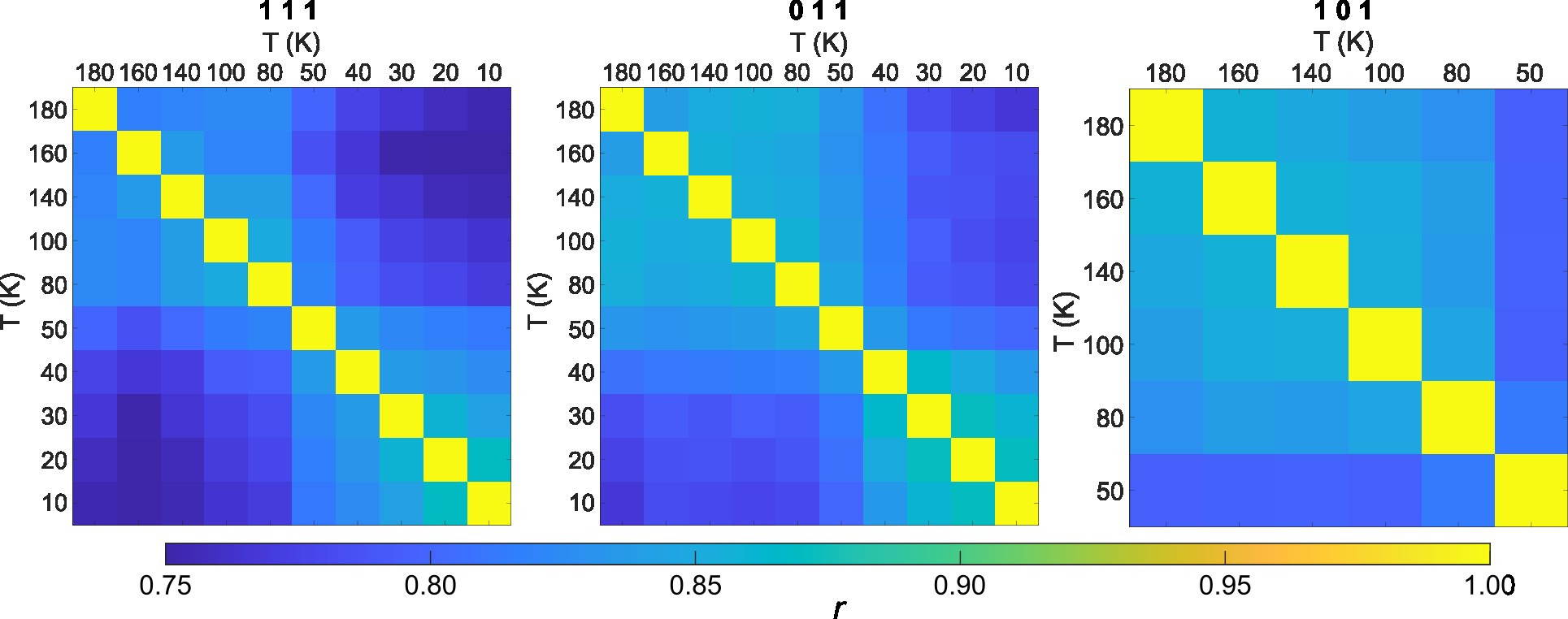}
    \caption{Pearson's $r$ coefficient heat maps of the Bragg peak intensity distribution (binarized) for different temperatures and reflections. The $101$ CXDP was not collected below 50 K.}
    \label{fig:CXDP_XC_B}
\end{figure}

\begin{figure}
    \centering
    \includegraphics[width=\linewidth]{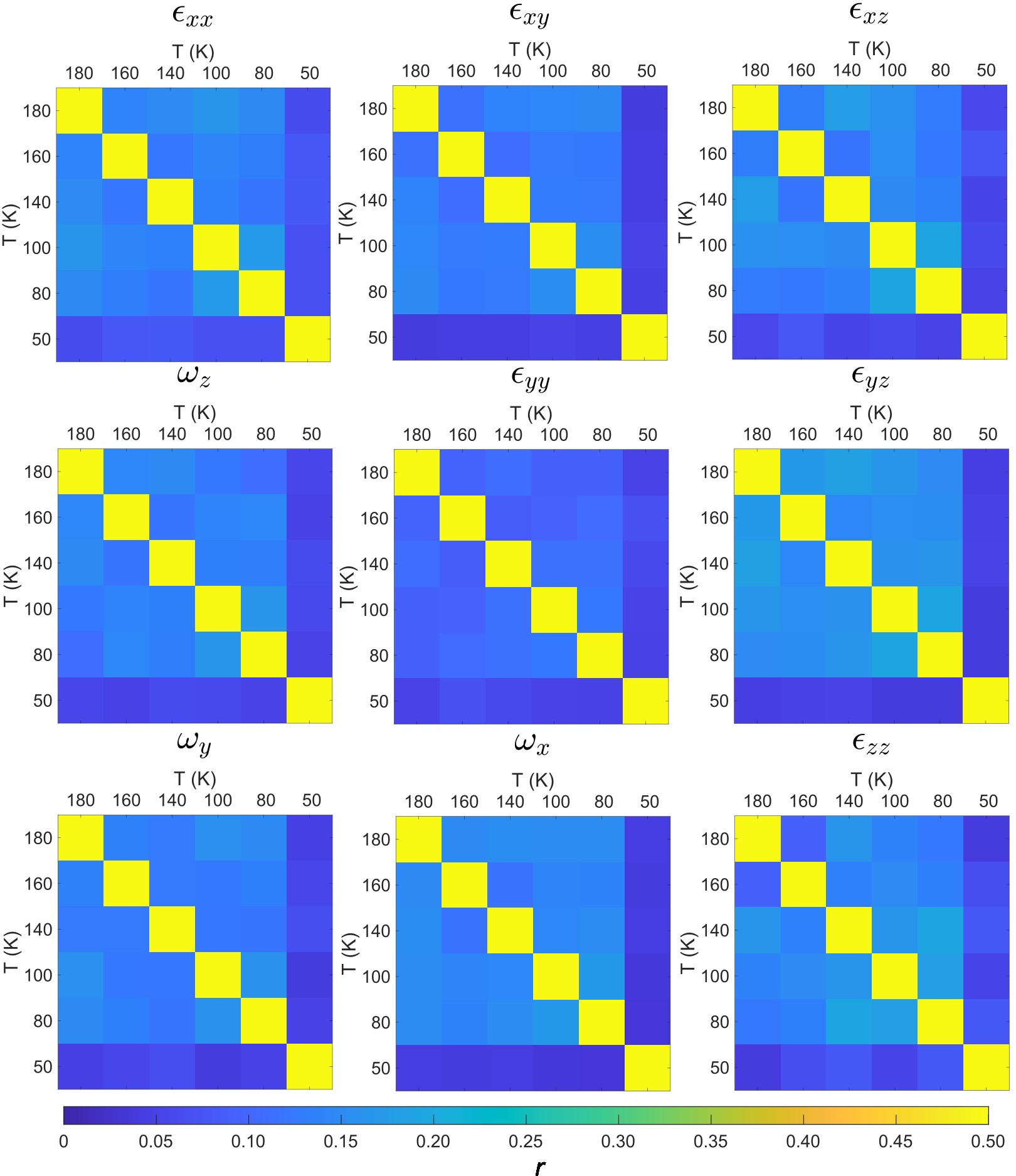}
    \caption{Pearson's $r$ coefficient heat maps of the tensor components for different temperatures.}
    \label{fig:Tensor_XC_B}
\end{figure}

Similar to the results presented in Sec. \ref{subsubsec:XC_A}, we employed Pearson's $r$ correlation coefficient heat maps analysis to identify changes in the CXDPs (Fig. \ref{fig:CXDP_XC_B}) and the tensor components (Fig. \ref{fig:Tensor_XC_B}).

\subsubsection{\label{subsubsec:Radial_tensor_B}Crystal B radial strain and rotation tensor magnitudes}
\begin{figure}
    \centering
    \includegraphics[width=\linewidth]{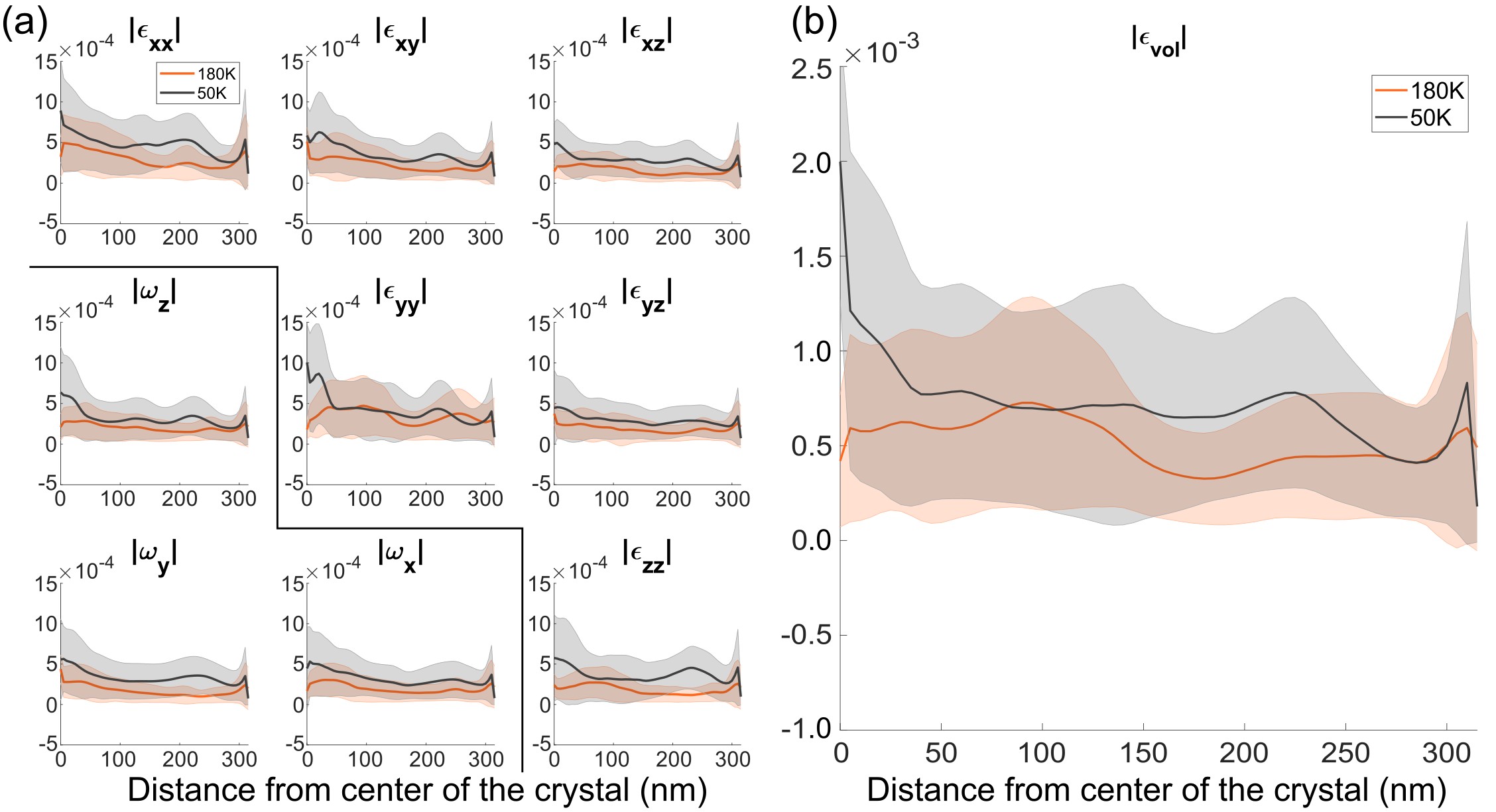}
    \caption{Histogram of the radial distribution of the tensor magnitudes for crystal B at 50 K (black) and 180 K (red). The shaded regions represent the standard deviation of the mean at each radial position from the center of the crystal. (a) shows the radial averages and spread of all the computed tensor components, while (b) shows the volumetric strain, \(\epsilon_{xx} + \epsilon_{yy} + \epsilon_{zz}\).}
    \label{fig:Radial_strain_B}
\end{figure}

The radial average magnitude of the tensor components for crystal B was compared at 50 and 180 K, as shown in Fig. \ref{fig:Radial_strain_B}. The radial histogram plots differ noticeably from those of crystal A, illustrating the different effects of strain due to dislocations and that of voids. Unlike in crystal A, both the volumetric strain and the tensor components are concentrated at the center of crystal B, seen both in increased magnitude and heterogeneity at small radius. Despite the differences in magnitude and distribution pattern, there is consistency with crystal A regarding the high- and low-temperature states of STO. The low-temperature state, on average, exhibits a higher magnitude and heterogeneity.

\section{\label{sec:Discussion}Discussion}
Based on MBCDI measurements of the two STO crystals, we observe the presence of defects in the form of dislocations in crystal A and STO impurities in crystal B. Both of these defects appear to be responsible for the increased magnitudes of local strain, spatially mapped out within the crystal by BCDI, notably due to the clear difference in their radial distributions. The strain fields generated by the dislocations in crystal A can penetrate throughout the entire crystal, but most of the strain field that is sensitive to BCDI may be masked. The noticeable level of STO impurities in crystal B also contributes to strain fields resembling inclusions, which are accentuated at lower temperatures, likely due to greater differences in thermal lattice expansion. With the 3D imaging capabilities of BCDI, we can isolate the defects from the remainder of each crystal to observe how a near-pristine crystal might behave.

There is a consistent transition in structure below 50 K for crystal A and crystal B, as shown in the heat maps in Figs. \ref{fig:CXDP_XC_A}, \ref{fig:Tensor_XC_A}, \ref{fig:CXDP_XC_B}, and \ref{fig:Tensor_XC_B}. There appears to be two dominant structural modes for crystal A: one below 50 and one above 50 K. Neither the heat maps for the CXDPs nor the tensor components show a sharp change between 50 and 100 K, but there is a consistent pattern across all heat maps, suggesting that these two states are present. Below 50 K is where low-temperature transitions start, as discussed in the introduction. 

\begin{figure}
    \centering
    \includegraphics[width=\linewidth]{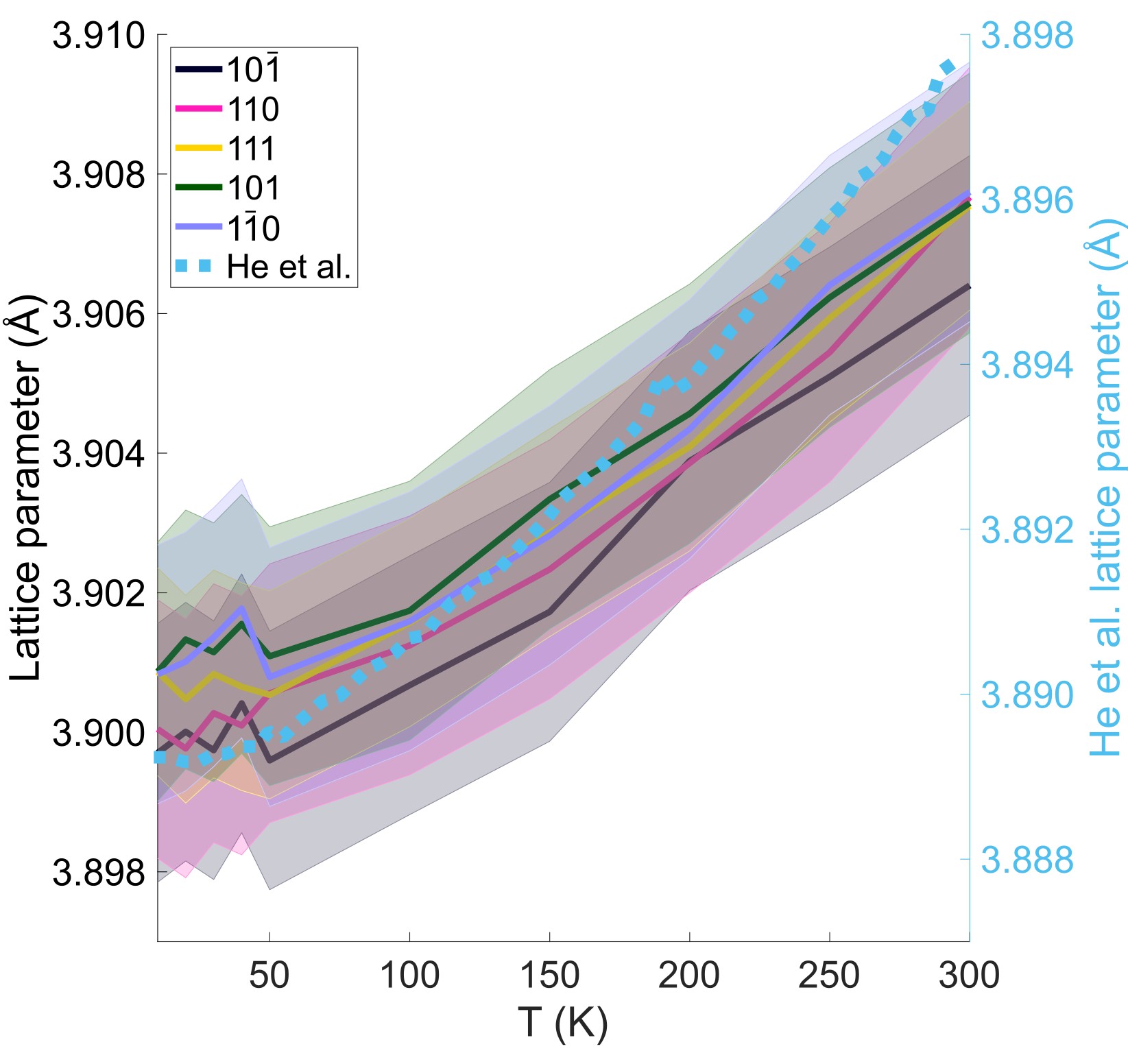}
    \caption{The lattice parameter for crystal A calculated based on the CXDP movement along the $2\theta$ direction as a function of temperature. The shaded region corresponds to the error associated with the sphere of confusion of the goniometer, and position of the sample along the beam with respect to the rotation center. The temperature trend agrees well with that of strained STO thin films from He \textit{et al.} \cite{He2003}.} % maybe little domains are forming below 50K, thus compensating for the lattice shrinking
    \label{fig:CXDP_positions}
\end{figure}

When we use the position of the Bragg peak along the $2\theta$ position to calculate the lattice parameter (Fig. \ref{fig:CXDP_positions}), we observe that the lattice parameter decreases linearly due to thermal contributions ($\sim10^{-5} \, \text{\AA}/\text{K}$) down to 50 K, where it then flattens out. This observation matches the lattice parameter trends on biaxially strained STO thin films reported by He \textit{et al.} \cite{He2003}, who reported no change in the lattice parameter as a result of the AFD transition, but noted that the lattice parameter remains relatively flat below 50 K. Loetzsch \textit{et al.} noted a similar lattice parameter trend for the $a$ axis near the surface of bulk STO \cite{Loetzsch2010}. This sub-50 K transition has also been noted for the thermal expansion coefficient, as reported by Tsunekawa \textit{et al.} \cite{Tsunekawa1984}. For crystal A, the coefficient of thermal expansion is $7.81 \times 10^{-6} \, \text{K}^{-1}$ between 100 and 300 K, and $1.24 \times 10^{-6} \, \text{K}^{-1}$ between 10 and 50 K. This could be due to damping of the polar soft modes reported below 50 K by Yamanaka \textit{et al.} \cite{Yamanaka2000}.

A possibility for the structural change below 50 K is ferroelectricity.  Ferroelectric behavior has been reported for strained STO from second-harmonic optical measurements \cite{Li2006}. Li \textit{et al.} reported the emergence of a ferroelectric phase transition in STO thin films depending on temperature and in-plane strain \cite{Li2006}. Based on thermodynamic phase models \cite{Li2006,Pertsev2000}, STO thin films can become ferroelectric at 0 K outside the strain range of $-2\times10^{-3}$ to $5\times10^{-4}$ (depending on Landau free energy coefficients), while considering the competing AFD phase. Although the strain fields here are not epitaxially induced, the strain tensor values, especially near dislocation cores and impurities ($>|3\times10^{-3}|$), exceed this threshold value. Although dislocations can result in other complex phenomena in STO, such as the interaction of dislocation cores with oxygen vacancies \cite{Jiang2011}, dislocations have been shown to stabilize flexoelectric polarization at dislocation cores experimentally and theoretically \cite{Zubko2007}. Strain gradients are known to break local centrosymmetry, which can lead to the stabilization of a ferroelectric state in STO. If the magnitude of the strain field exceeds the threshold required for a ferroelectric state \cite{Li2006,Pertsev2000}, the strain fields induced by impurities, in the case of crystal B, could also stabilize local polarization.

Another possibility for stabilization is the surface strain caused by the size of the microcrystals compared to the bulk. The atoms on the surface of the crystals experience a higher surface energy relative to the bulk. To relieve the surface energy, the atoms can distort themselves, thereby creating surface strain. Reconstructions of the microcrystals shown in Fig. \ref{fig:Dipole}, and figures \ref{fig:Individual_reconstructions_A}, \ref{fig:Individual_reconstructions_B}, and \ref{fig:111_reconstruction} in the appendices have regions of lattice expansion (yellow) on the surface, which lead to surface strains Fig. \ref{fig:Radial_strain_A} and \ref{fig:Radial_strain_B}. The surface tension on these microcrystals could be analogous to the epitaxial strain induced on STO thin films \cite{Loetzsch2010}, suggesting that the ferroelectric state of strained STO thin films \cite{He2003, Li2006, Haeni2004} might also be relevant for these microcrystals.

Although the ferroelectric polarization state has been proposed to be along the $[0\;0\; 1]$ direction under compressive biaxial strain, and either along a $<0\;0\;1>$ or $<1\;1\;0>$ in-plane direction under tensile strain \cite{Li2006}, the polarization state cannot be observed in the BCDI reconstructions. The ferroelectric phase will induce a polarization that changes the lattice parameter slightly, much like the AFD transition, but here we do not observe any split Bragg peaks. Ferroelectric polarization has been probed with BCDI \cite{Diao2024}, but the image of the ferroelectric state is not clear due to the dominant strain fields induced by the defects.

We do not observe any noticeable changes in the heat maps (Figs. \ref{fig:CXDP_XC_A}, \ref{fig:Tensor_XC_A}, \ref{fig:CXDP_XC_B}, and \ref{fig:Tensor_XC_B}) corresponding to the cubic-to-tetragonal AFD phase transition at 105 K, nor do we observe a change in lattice parameter or change of slope (thermal expansion coefficient) in Fig. \ref{fig:CXDP_positions}. Interestingly, He \textit{et al.} reported that the AFD transition can occur as much as 50 K higher for biaxially strained STO thin films compared to bulk STO \cite{He2003}. Their experiment measured a half-order superlattice Bragg peak $(\frac{1}{2},\frac{1}{2},\frac{7}{2})$ associated with TiO$_{6}$ octahedra rotation, and they observed a gradual decrease in intensity across different film thicknesses. Hence, it could be that the AFD phase transition occurs in crystal A and crystal B, but it cannot be detected from our measured crystal Bragg peaks or the lattice parameter change in Fig. \ref{fig:CXDP_positions}. Furthermore, the rotation of the TiO$_{6}$ octahedra around the $[1\;0\;0]$ crystallographic axis and the $c/a$ ratio for the tetragonal phase is below the strain resolution for BCDI. However, the AFD phase transition could also be influenced by the surface area-to-volume ratio and the chemistry of the microcrystals, which are not explored here.

\begin{comment}
   At low temperatures, STO exhibits quantum paraelectric behavior, where quantum fluctuations prevent the system from fully transitioning into a ferroelectric phase. Strain induced by defects can either enhance or disrupt these quantum fluctuations by introducing localized polarization effects that can interact with the intrinsic dipole moments in the material. For instance, defects can pin ferroelectric domains, preventing them from switching under an external electric field. This pinning effect can modify the dielectric response of the material, which could be an explanation for the high dielectric constants that deviate from the classical Curie-Weiss law below 50 K \cite{Weaver1959}.  
\end{comment}

\section{\label{sec:Conclusion}Conclusion}
This study successfully leverages the capability of MBCDI to resolve the residual strain and rotation tensor components for two STO microcrystals as they are cooled down to cryogenic temperatures. Below 50 K, each microcrystal transitioned to its low-temperature state, exhibiting increased strain magnitude and heterogeneity throughout the entire crystal. We do not observe the AFD phase transition in the MBCDI analysis, even though it should be present in strained samples \cite{He2003}; however, it might be suppressed by a ferroelectric phase \cite{Zhong1995,Yamanaka2000}. Another possibility is the size or chemistry dependence of the hydrothermally synthesized STO that alters the phase transition behavior. Our results show structural modes in STO microcrystals at low temperatures, revealing information about structural transitions at the nanoscale. 

\section{\label{sec:Data availability}Data availability}
The processed diffraction patterns, final reconstructions, and data analysis scripts are publicly available at \cite{zenodo_ref}.

\begin{acknowledgments}
We would like to thank Steve Shapiro for enlightening us about the neutron scattering work on STO, and Felix Hofmann for thoughtful discussions about dislocation analysis. Work at Brookhaven National Laboratory was supported by the U.S. Department of Energy, Ofﬁce of Science, Ofﬁce of Basic Energy Sciences, under Contract No. DESC0012704. The authors acknowledge Diamond Light Source for time on Beamline I16 under proposal number MM30687, MM33417, and MM34617. N.Z. and Z.A. acknowledges support from the NSFC (Grant No. 12161141012) S. H., S. C., H. N., M. N., and H. K. acknowledges support from the National Research Foundation of Korea (Grant NRF-2021R1A3B1077076). Work performed at UCL was supported by EPSRC.

%burgers vector in the plane is shear loop
\end{acknowledgments}

\appendix
\section{\label{appendix:phase_retrieval}Phase retrieval} 
The reconstruction process for each CXDP was performed independently for each reflection. The cropped CXDP on the detector was padded with zeros to a size of $256 \times 256 \times 128$ pixels and binned by a factor of 2 along the detector plane, resulting in a data size of $128 \times 128 \times 128$ pixels.

Each reconstruction was initialized with a random guess. A guided phasing approach \cite{Chen2007} with 60 (50 for the crystal shown in Fig. \ref{fig:Dipole}) individuals and 4 generations was used, employing a geometric average breeding mode. For each generation and population, a block of 20 error reduction (ER) and 180 hybrid input-output (HIO) iterations, with $\beta = 0.9$, was repeated three times. This was followed by 20 ER iterations to finalize the object. The shrinkwrap algorithm \cite{Marchesini2003} with a threshold of 0.1 was used to update the real-space support at each iteration. The $\sigma$ values for the Gaussian kernel shrinkwrap smoothing for each generation were $\sigma = 2.0, 1.5, 1.0, 1.0$, respectively. The best reconstruction was determined using a sharpness criterion, as it is an appropriate metric for crystals containing defects \cite{Ulvestad2017b}. From the 30 (25 for the crystal shown in Fig. \ref{fig:Dipole}) best reconstructions, candidates well-correlated to the most similar candidate were averaged to produce the final reconstruction. This process was repeated again, with the reconstructions from the first round now as starting guesses.

The overall average 3D spatial resolution was 63.7 nm for crystal A and 61.0 nm for crystal B. This was determined by differentiating the electron density amplitude across the crystal/environment interface for five linearly independent directions and fitting a Gaussian to each of the profiles. The reported spatial resolution was the average full width at half maximum of the Gaussian profiles, across all measured reflections and temperatures for each crystal.

%spatial resolutions for each temperature: 56.11 + 58.8 + 58.73 + 69.05 + 67.86 + 70.36 + 60.72 + 72.81 + 57.97 + 64.1
%spatial resolutions for each temperature: 59.72 + 60.88 + 55.48 + 56.64 + 64.14 + 74.44

\section{\label{appendix:Individual_reconstructions}Individual reconstructions} 
\begin{figure}
    \centering
    \includegraphics[width=\linewidth]{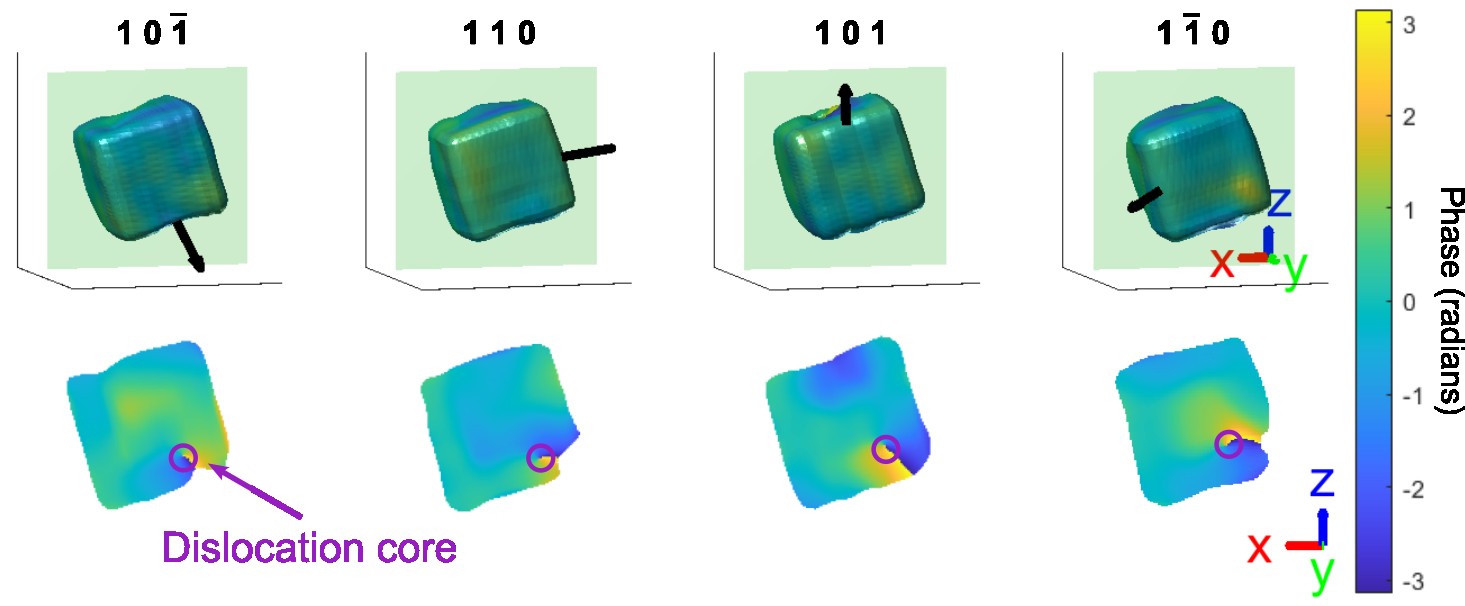}
    \caption{The top row shows the morphology of the $10\bar{1}$, $110$, $101$, and $1\bar{1}0$ reconstructions for STO crystal A at 300 K, colored by phase. The normalized amplitude threshold is 0.20. The black arrow indicates the direction of the measured scattering vector $\mathbf{Q_\mathit{hkl}}$. The bottom row shows slices through the phase at $y = 2.5$ nm (green plane shown in the top row). The dislocation core positions, characterized by a phase vortex, are within the purple circle. The coordinate axes arrows have a length of 100 nm.}
    \label{fig:Individual_reconstructions_A}
\end{figure}

The individual reconstructions for crystal A are shown in Fig. \ref{fig:Individual_reconstructions_A}. Here, the phase vortices, shown within the purple circles, represented the positions of the dislocation cores \cite{Clark2015} and were used to determine the dislocation lines discussed further in Appendix \ref{appendix:Dislocation_analysis}. Fig. \ref{fig:Individual_reconstructions_B} shows the individual reconstructions for crystal B.

\begin{figure}
    \centering
    \includegraphics[width=\linewidth]{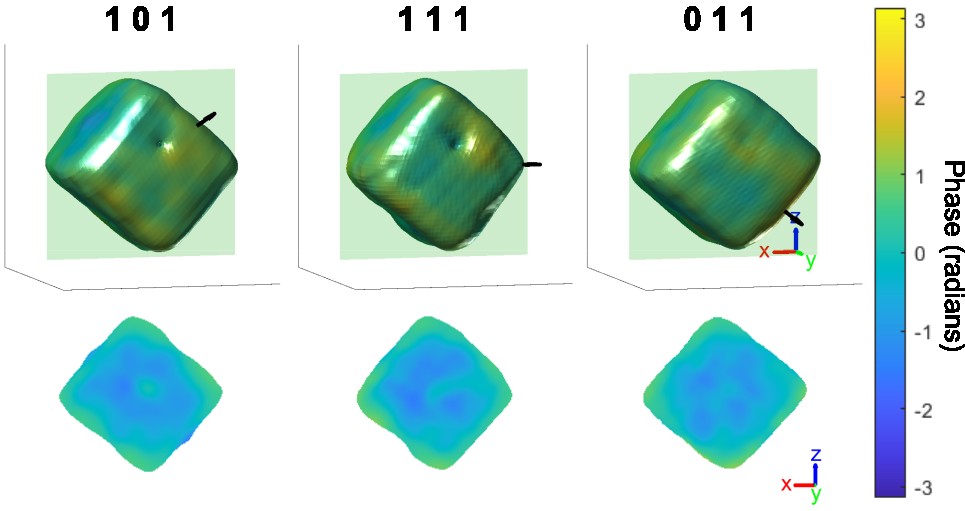}
    \caption{The top row shows the morphology of the $101$, $111$, and $011$ reconstructions for STO crystal B at 180 K, colored by phase. The normalized amplitude threshold is 0.20. The black arrow indicates the direction of the measured scattering vector $\mathbf{Q_\mathit{hkl}}$. The bottom row shows slices through the phase at $y = 2.5$ nm (green plane shown in the top row). The coordinate axes arrows have a length of 100 nm.}
    \label{fig:Individual_reconstructions_B}
\end{figure}

\begin{figure}
    \centering
    \includegraphics[width=\linewidth]{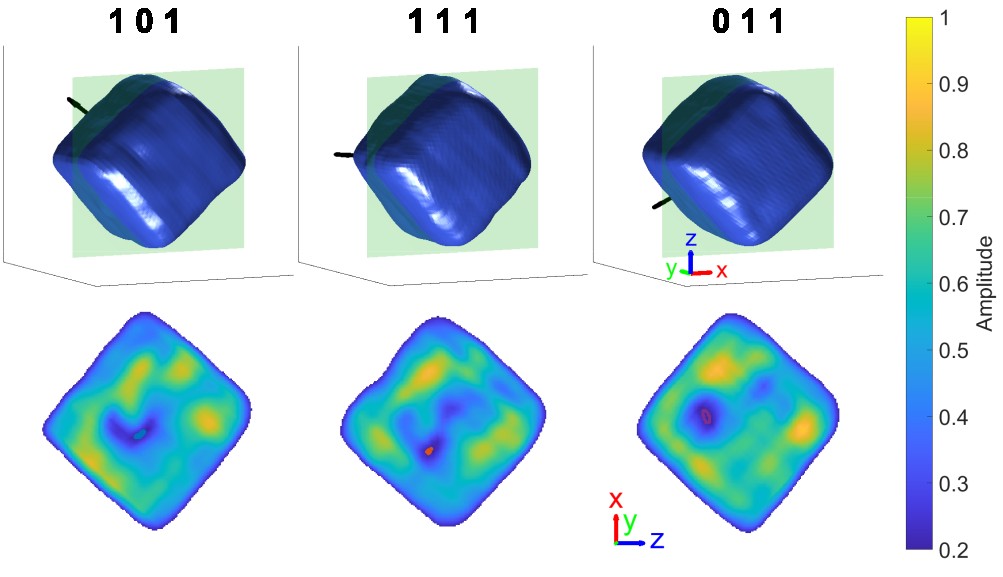}
    \caption{The top row shows the morphology of the $101$, $111$, and $011$ reconstructions for STO crystal B at 180 K, colored by amplitude. The normalized amplitude threshold is 0.20. The black arrow indicates the direction of the measured scattering vector $\mathbf{Q_\mathit{hkl}}$. The bottom row shows slices through the amplitude at $y = -92.5$ nm (green plane shown in the top row). The small rings outline the voids in each reflection, as in Fig. \ref{fig:Tensor_slices_B}. The coordinate axes arrows have a length of 100 nm.}
    \label{fig:Individual_reconstructions_B_amp}
\end{figure}

\section{\label{appendix:Dislocation_analysis}Dislocation analysis}
\begin{figure}
    \centering
    \includegraphics[width=\linewidth]{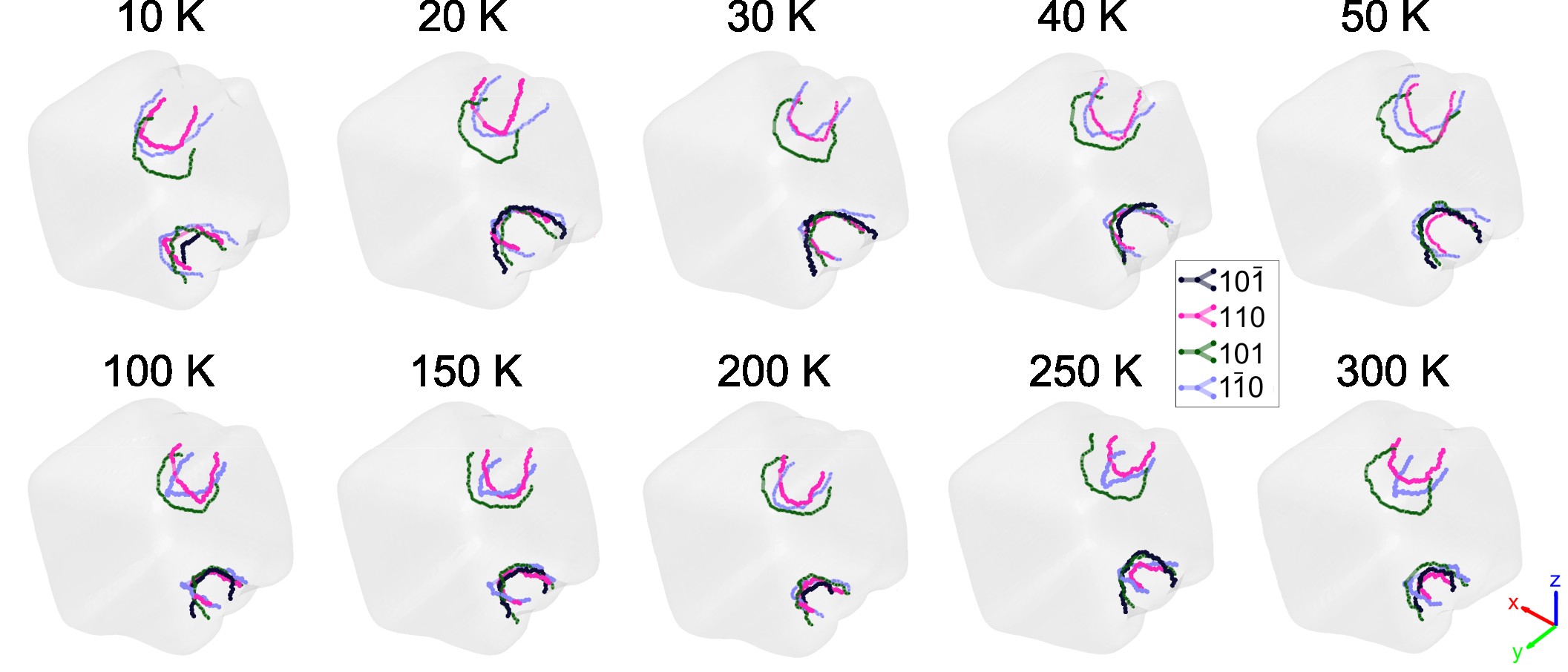}
    \caption{The dislocation lines for each reflection are plotted against the average amplitude for crystal A across all measured temperatures. Note that not all the lines terminate at the surface of the average amplitude, but they do for their respective amplitudes. The bottom dislocation is used to align amplitudes and phases in Sec. \ref{subsec:Strain_and_rotation_tensors}. These dislocation lines are averaged to provide the dislocation positions shown in Fig. \ref{fig:Dislocation_lines_avg}. The coordinate axes arrows have a length of 100 nm. The normalized average amplitude threshold is 0.20.}
    \label{fig:Dislocation_lines_all}
\end{figure}

We define the dislocation line as the locus of low-amplitude voxels lying along its core, referred to as nodes. With limited resolution, these correspond to cancellations of unresolved regions of the phase vortex surrounding the dislocation core. In crystal A, each dislocation is composed of many nodes (e.g., Fig. \ref{fig:Individual_reconstructions_A}), joined by edges based on the MATLAB graph object. The nodes were determined automatically by integrating the derivatives of the complex exponential of the phase ($e^{i\psi_{hkl}(\mathbf{r})}$) \cite{Yang2022c} and then selecting the maximum value as the dislocation node position if the value exceeded a threshold. If the resulting dislocation lines were fragmented inside the crystal due to a missing node, they were joined together to create a continuous line so that they did not terminate inside the crystal. The dislocations were found not to overlap perfectly in the reconstructions from different Bragg peaks, which we attribute to noise propagation. To increase overlap, the phases of each reflection were shifted such that the resulting dislocation lines best overlapped with the bottom dislocation of the $10\bar{1}$ reflection. The resulting dislocation lines for each reflection are shown in Fig. \ref{fig:Dislocation_lines_all}.

Dislocations are visible only when $\mathbf{Q_\mathit{hkl}} \cdot \mathbf{b} \neq 0$ \cite{Williams2009} and are therefore not necessarily visible for all reflections. For instance, the top dislocation did not appear in the $10\bar{1}$ reflection but was visible in the others, as shown in Fig. \ref{fig:Dislocation_lines_all}. We also noted that the dislocation lines lie on the $\{\bar{1}\;1\;0\}$ plane family, suggesting that the Burgers vectors are likely of the type $a<1\;1\;0>$ \cite{Klomp2023}. To identify the Burgers vectors for the dislocations, we computed $\mathbf{Q_\mathit{hkl}} \cdot \mathbf{b}$ for the possible Burgers vectors in Table \ref{table:Q_dot_b}. However, we cannot determine the sign of the Burgers vector.

\begin{table}
    \caption{$\mathbf{Q_\mathit{hkl}}\mathbf{\cdot b}$ dislocation visibility}
    \begin{center}
        \begin{tabular}{c|ccccc}      % Alignment for each cell: l=left, c=center, r=right
            \multicolumn{1}{c|}{$\mathbf{b}\symbol{92}\mathbf{Q_\mathit{hkl}}$} & $10\bar{1}$ & $110$ & $101$ & $1\bar{1}0$ & $111$ \\ \hline
            $a[1\;1\;0]$ & visible & visible & visible & - & visible\\
            $a[\bar{1}\;1\;0]$ & visible & - & visible & visible & -\\
            $a[1\;0\;1]$ & - & visible & visible & visible & visible\\
            $a[\bar{1}\;0\;1]$ & visible & visible & - & visible & -\\
            $a[0\;1\;1]$ & visible & visible & visible & visible & visible\\
            $a[0\;\bar{1}\;1]$ & visible & visible & visible & visible & -\\
        \end{tabular}
        \label{table:Q_dot_b}
    \end{center}
\end{table}

\begin{figure}
    \centering
    \includegraphics[width=\linewidth]{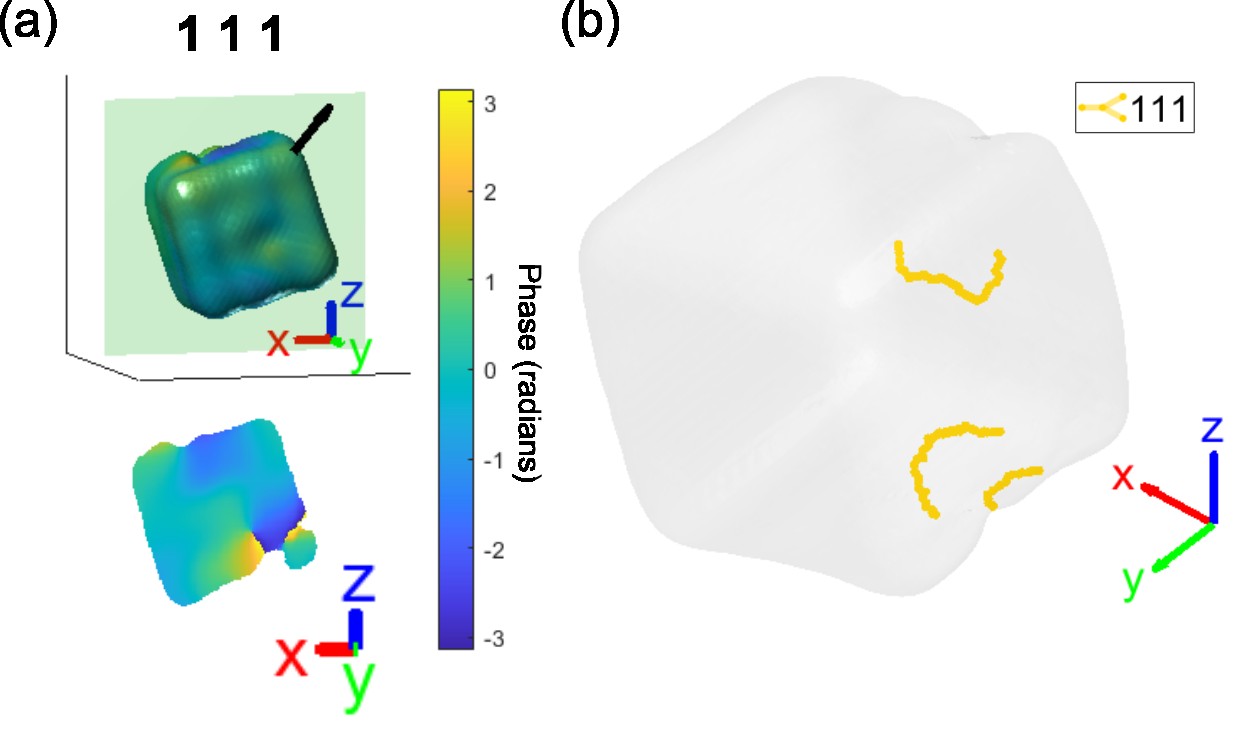}
    \caption{The $111$ reconstruction at 300 K that is not used in the tensor computation due to poorer reconstruction quality at different temperatures. The top row of (a) shows the morphology, colored by phase. The bottom row shows a slice through the phase at $y = 2.5$ nm (green plane shown in the top row). (b) shows the dislocation lines plotted against the average amplitude for crystal A. For the entire figure, the coordinate axes arrows have a length of 100 nm and the normalized average amplitude threshold is 0.20.}
    \label{fig:111_reconstruction}
\end{figure}

The $111$ reconstruction is shown in Fig. \ref{fig:111_reconstruction}. Note that there was a third dislocation present at the bottom that was not consistently present in the other reflections.

% The \nocite command causes all entries in a bibliography to be printed out
% whether or not they are actually referenced in the text. This is appropriate
% for the sample file to show the different styles of references, but authors
% most likely will not want to use it.
%\nocite{*}

\bibliography{library}% Produces the bibliography via BibTeX.

\end{document}